\documentclass[journal]{IEEEtran}
\usepackage{lscape}
\usepackage{adjustbox}
\usepackage{rotating} %
\usepackage{float}
\usepackage{afterpage}
\usepackage{amsmath}
\usepackage{amssymb}
\usepackage{booktabs}
\usepackage{mathtools}
\usepackage{graphicx} %

\makeatletter
\let\MYcaption\@makecaption  %
\makeatother

\usepackage{subcaption}
\makeatletter
\let\@makecaption\MYcaption  %
\makeatother

\usepackage{tikz}
\usepackage{siunitx}

\usepackage{url}
\usepackage{multirow}
\usepackage{titlesec}
\usepackage{hyperref}
\hypersetup{
	colorlinks=true,
	linkcolor=black,
	urlcolor=black,
	citecolor=black
}

\usepackage{xcolor}

\usepackage{pifont}
\newcommand{\cmark}{\checkmark}
\newcommand{\xmark}{\ding{55}} %

\newtheorem{definition}{Definition}

\titlespacing*{\section}{0pt}{0.2\baselineskip}{0.2\baselineskip}
\titlespacing*{\subsection}{0pt}{0.2\baselineskip}{0.2\baselineskip}

\newcommand{\dom}{{\mbox{\bf dom}}}

\allowdisplaybreaks

\newcommand{\minlp}{\texttt{MINLP}}
\newcommand{\nonsmooth}{\texttt{nonsmooth}}

\newcommand{\softplus}{\texttt{softplus}}
\newcommand{\spline}{\texttt{spline}}

\newcommand{\smallnw}{\texttt{30-bus}}
\newcommand{\mediumnw}{\texttt{99-bus}}
\newcommand{\largenw}{\texttt{539-bus}}

\begin{document}

\title{Volt-VAr-Watt Optimization in Four-Wire Low-Voltage Networks: Exact Nonlinear Models and Smooth Approximations - Extended Version}
\author{Sleiman~Mhanna,~\IEEEmembership{Senior Member,~IEEE,}
	Frederik~Geth,
	Lucas~Quiertant,~\IEEEmembership{Member,~IEEE,}
	and~Pierluigi~Mancarella,~\IEEEmembership{Fellow,~IEEE}%

	\thanks{S. Mhanna, L. Quiertant, and P Mancarella are with the Electrical and Electronic Engineering Department, The University of Melbourne, Melbourne, Victoria 3010, Australia.
		F. Geth is with the School of Electrical Engineering and Computer Science, The University of Queensland, QLD, Australia.
		Corresponding author: sleiman.mhanna@unimelb.edu.au.}%
		
	\thanks{This manuscript is an extended version of S. Mhanna, L. Quiertant, F. Geth, and P. Mancarella, “Volt-VAr-Watt Optimization in Four-Wire Low-Voltage Networks: Exact Nonlinear Models and Smooth Approximations,” IEEE Trans. Power Syst., vol. 41, no. 5, pp. 3663–3677, 2026, doi: 10.1109/TPWRS.2026.3677246.}
	\thanks{This work was supported in part by the EPICS Global Center.}
}

\markboth{}{}
\maketitle

\begin{abstract}
	The proliferation of distributed energy resources is increasing the prevalence of both overvoltages and undervoltages in low-voltage (LV) distribution networks.
	Smart inverter functionalities, such as Volt-VAr and Volt-Watt control, can regulate voltage at the consumer level but are challenging to capture in optimization models due to their \textit{nondifferentiability}.
	This paper proposes three nonlinear models for four-wire unbalanced optimal power flow that incorporate these nonsmooth functions without binary or integer variables.
	The first model encodes these nonsmooth functions directly as user-defined functions using control flow, a feat enabled by many state-of-the-art algebraic modeling languages.
	The second and third introduce bespoke \textit{smooth} approximations with tunable approximation errors to address potential numerical issues arising from nondifferentiability, improving reliability while maintaining accuracy.
	All three methods are evaluated on real four-wire unbalanced LV networks with varying rooftop solar adoption levels, including a 539-bus system with 302 smart inverters.
	This paper is not only the first to demonstrate accurate, reliable, and tractable Volt-Var-Watt optimization (VVWO) on real four-wire unbalanced LV network models, but it also establishes that optimization-based, non-incremental methods offer superior reliability compared to commonly used incremental (quasi-steady-state) approaches.
\end{abstract}

\begin{IEEEkeywords}
	Volt-VAr control, Volt-Watt control, four-wire low-voltage network modeling, nonlinear optimization, smooth approximations, softplus, splines, unbalanced optimal power flow.
\end{IEEEkeywords}

\setlength{\belowdisplayskip}{0.5pt} \setlength{\belowdisplayshortskip}{0.5pt}
\setlength{\abovedisplayskip}{0.5pt} \setlength{\abovedisplayshortskip}{0.5pt}
\setlength{\textfloatsep}{1pt plus 1.0pt minus 1.0pt}
\setlength{\floatsep}{1pt plus 1.0pt minus 1.0pt}

\section{Introduction}\label{sec_introduction}

\subsection{Challenges in Power Distribution Systems}
\IEEEPARstart{T}{he} widespread adoption of behind-the-meter distributed energy resources (DERs), such as rooftop photovoltaic (PV) panels, battery energy storage systems (BESS), and electric vehicles (EVs), is creating unprecedented operational challenges for electricity distribution networks (DNs).
These challenges include increased frequency and severity of overvoltages, undervoltages, voltage unbalance, as well as accelerated degradation of overhead lines and transformers due to frequent overloading (i.e., thermal limit violations).
Rooftop PV systems can cause severe overvoltages, especially during peak generation periods when local production exceeds consumption.
Conversely, uncoordinated charging of EVs can lead to substantial undervoltages, particularly when large numbers of EVs are charging simultaneously.
Moreover, the uneven distribution of DERs across phases exacerbates voltage unbalance, which is observed in LV networks typically due to unbalanced loads and untransposed lines \cite{kersting2017distribution}.
This unbalance can result in higher network losses emanating from a rise in neutral current, which in some cases may exceed phase currents \cite{BARBATO201876}.

Distribution utilities traditionally manage voltages through devices such as tap changers, voltage regulators, or capacitor banks.
However, these devices are typically located on the primary side of the MV/LV distribution feeder, and their taps are adjusted only a few times per year.
In general, distribution utilities lack visibility on voltage fluctuations on the LV side of the feeder, and traditional devices are ill-equipped to respond to fast and local voltage fluctuations emanating from the increasing uptake in EVs and rooftop PV \cite{Mahmud2016}.
To alleviate this issue, smart inverter (SI) functionalities such as droop-based Volt-VAr control (VVC) and Volt-Watt control (VWC) have recently been introduced to support voltage regulation at much shorter timescales than traditional equipment.
VWC can also help reduce overloading through active power curtailment.
Being located behind the meter, and therefore fully agnostic to network topology, they are often thought of as \textit{decentralized} approaches to voltage regulation \cite{Jabr2019_RobustVVC}.
Nonetheless, Lusis et al. \cite{8810940} highlight that such a passive and decentralized voltage regulation quickly reaches its limits in LV networks with high integration of PV.
Subsequent studies by Mohanan et al. \cite{Mohanan_2020} and Lusis et al. \cite{9222371} demonstrate that \textit{coordinated} SI dispatch could further enhance voltage regulation in unbalanced LV DNs.

Such coordination can be modelled as a distribution optimal power flow (OPF) problem with Volt-VAr-Watt optimization (VVWO) that can optimally dispatch active and reactive power from DERs.
Farivar et al. \cite{farivar2011optimal} and Alboaouh et al. \cite{MISQ_VVWO} proposed solving this VVWO problem by assuming that VVWO can be obtained directly from the PQ capability curve of the PV inverter.
However, modern inverter VVC and VWC are often programmed to adhere to droop settings, as defined in IEEE-1547-2018 \cite{IEEE_Standard_1547}, rendering the inverter setpoints obtained in \cite{farivar2011optimal,MISQ_VVWO} inaccurate and infeasible in practice.

Interestingly, strict adherence to the droop setting stipulated in IEEE-1547 \cite{IEEE_Standard_1547} is a rather rare occurrence in practice.
Recent field testing of these grid-supporting characteristics under rapidly changing voltages shows that, in real-world applications, voltages almost never lie precisely on the prescribed curves (and have somewhat significant VVWC response times on the order of $\sim$10s \cite{Tafti2021}), but instead lie within a band of $\pm 7.5\%$ of the target value \cite{csiro_report_test_package}.
In fact, distribution power flow (PF) simulation tools like \textsc{OpenDSS} \cite{Dugan2016_OpenDSS} (by default) model VVWC by allowing a small deviation from the idealized VVWC settings, and treat VVWC functions in a separate control module with a different tolerance to that of the PF subroutines \cite{Radatz2021_OpenDSSwithDER}.

\subsection{Research gaps: PF and VVWC}

\begin{figure}[t]
	\centering
	\includegraphics[width=1\columnwidth]{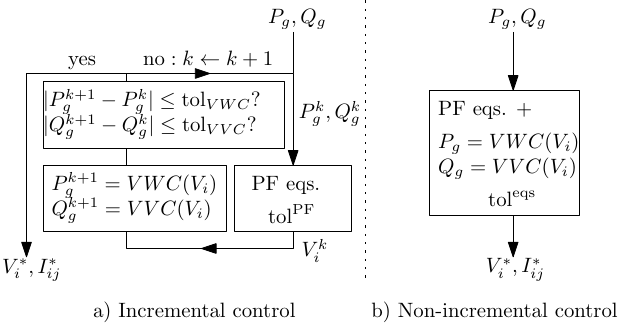}
	\caption{Comparison of incremental (a) vs non-incremental (b) control methods for solving the power flow with VVWC. Note that the incremental strategy requires additional tolerances for the VVWC modules.}
	\label{fig:loops}
\end{figure}

Methods that decouple control and network calculations, commonly used in distribution system simulation tools such as OpenDSS \cite{Dugan2016_OpenDSS,Radatz2021_OpenDSSwithDER}, fall under a general class of methods, hereafter referred to as \textit{incremental} methods \cite{Zhou2018_ReverseandForwardEngineeringforVVC,Eggli2020_StabillityofVVC}.
Their general architecture is illustrated in Fig.~\ref{fig:loops}a.
This type of method alternates between two separate modules: (i) a steady-state PF which computes bus voltages and line currents based on the current operating point, and (ii) the inverter control module which updates reactive power, VVC, and VWC setpoints based on the most recent voltages.
In this incremental structure, inverter set-points are updated based on the voltages obtained from the previous PF solution, after which a new PF is solved.
This sequential process repeats until both voltages and inverter outputs settle to a consistent operating point. This process can also be described as an approximation of the time-domain control loop via step discretization, i.e., \textit{quasi steady-state} simulation.

Whereas OpenDSS solves a PF in the inner layer \cite{Radatz2021_OpenDSSwithDER}, the incremental VVC approach in \cite{Colot2024_RobustIncrementalVVC} solves a (robust) OPF, which can enforce voltage limits and other variable bounds.
However, unlike OpenDSS, which can solve PF in unbalanced four-wire network to exactness, the method in \cite{Colot2024_RobustIncrementalVVC} linearizes the PF equations and adopts a ``single-phase equivalent'' representation of the DN.
In general, the decoupling of the control module and the PF module in incremental approaches simplifies the implementation but this alternating process can lead to slow convergence or oscillatory behavior, particularly when the network is highly unbalanced, when neutral currents are significant, or when VVC and VWC curves are steep.\footnote{This is demonstrated numerically in Section~\ref{sec_numericaleval}.}

In contrast, \textit{non-incremental} control strategies, illustrated in Fig.~\ref{fig:loops}b, solve the PF and inverter control steps simultaneously, in a single, \textit{unified set of equations} rather than through sequential updates.
Non-incremental methods therefore eliminate the outer iteration by embedding the control equations directly into the network equations.
This can produce a more consistent solution for voltages and inverter set-points and can improve reliability in LV feeders where strong coupling between phases and between active and reactive power can make the iterative process in incremental methods unreliable.

Despite these potential advantages, non-incremental control methods are vulnerable to an intrinsic property of VVC and VWC functions, namely their \emph{piecewise-linear} (PWL) nature.
In general, PWL functions introduce complexity into Newton-based PF solvers because they are not differentiable at their breakpoints.
Newton-type methods rely on smooth, differentiable functions to construct the Jacobian and update iterates.
When a control function switches from one linear segment to another, the associated derivative can undergo an abrupt change, which in turn translates into abrupt and discontinuous changes in the Jacobian.
These discontinuities can lead to poor local linearization, oscillatory behavior, or divergence, particularly when operating points lie near a breakpoint.
Nonetheless, a number of approaches have been proposed in recent literature to tackle this potential vulnerability.
In \cite{Godoy2024_VVC}, which embeds VVC directly into the Newton-Raphson PF solver, convergence is improved through a heuristic that dynamically limits reactive power by changing bus types in an alternating fashion between PV and PQ.
In contrast, \cite{TurnerBandele2021_SplineforVVCPF} approximates the PWL VVC function by a smooth function in the form of a cubic spline, making it  conducive for Newton-type PF solvers.

Although all the approaches discussed so far can be applied to four-wire unbalanced networks with explicit neutrals (that is, without Kron reduction), their reliability has not been thoroughly quantified.
Moreover, the non-incremental methods in \cite{Godoy2024_VVC,TurnerBandele2021_SplineforVVCPF} are restricted to PF applications rather than OPF formulations.
In contrast, optimization-based methods unlock a much broader range of capabilities, including dynamic operating envelopes, multi-period DER coordination, and distribution network planning under \textit{bespoke} constraints.

\subsection{Research gaps: OPF and VVWO}

Unsurprisingly, optimization-based approaches involving VVWO, which fall under the class of non-incremental control strategies in Fig.~\ref{fig:loops}b but with an OPF instead of a PF, face several challenges introduced by the PWL nature of VVWC.
One way to address the challenges introduced by the nonsmooth nature of these PWL functions is to encode them with binary or integer variables, transforming the problem into a mixed-integer nonlinear programming (MINLP) problem that can be solved with nonlinear branch-and-bound (NLBB) methods available in \textsc{Knitro} \cite{KNITRO} and \textsc{Juniper} \cite{juniper}.
Unfortunately, this class of MINLP problems, which have a nonconvex continuous relaxation,\footnote{The continuous relaxation is obtained by relaxing the integrality constraints to continuous ones.}
quickly become intractable, as demonstrated in Section~\ref{sec_numericaleval}, and in \cite{Dimoulias2025_DroopAidedSE} and \cite{Gebbran2023_three_block_ADMM} for state-estimation and distributed VVO in balanced MV DNs.
To bestow tractability, existing methods often resort to a combination of three strategies, \textit{convex relaxations}, \textit{approximations}, and \textit{simplifications} of the PF in unbalanced DNs.
A convex second-order cone relaxation (SOCR), also known as the DistFlow model,\footnote{The DistFlow model was originally introduced in \cite{Baran1989_DistFlow}.}, is adopted in \cite{Savasci2021_SOCP_VVC} to reduce the problem to a tractable mixed-integer second-order cone program (MISOCP).
Computational efficiency can be further improved with the help of approximations, such as those in \cite{Inaolaji2021_LinDist3Flow_VVWO,Soltani2023_VVC,Inaolaji2023_DOPFwithVVWOLinDist3Flow}, which rely predominantly on some form of \textit{linearization} of the OPF equations.
Typical linearization choices are the LinDist3Flow formulation as in \cite{Inaolaji2021_LinDist3Flow_VVWO,Inaolaji2023_DOPFwithVVWOLinDist3Flow},\footnote{The LinDist3Flow was first introduced in \cite{Arnold2016_LinDist3Flow}.}
or a Taylor series approximation around an operating point as in \cite{Soltani2023_VVC,Russell2025_DesignofVVC}.
This linearization, combined with the binary encoding of the VVC and VWC functions, further reduces the problem to a more tractable mixed-integer linear program (MILP).

Unfortunately, this improvement in computational efficiency comes at the expense of accuracy.
Since they do not contain the original nonconvex feasible set, approximations yield physically infeasible solutions, even in the absence of VVWO.
On the other hand, despite containing the original nonconvex feasible set, convex relaxations of the PF only yield physically feasible solutions under specific conditions that are rarely satisfied in practice \cite{Usman2020_SDPFourWireDPF}.

Furthermore, all of the existing optimization-based methods discussed so far adopt some form of simplification to the physics of PF.
The methods in \cite{Olowu2021_VVC,Savasci2021_SOCP_VVC,Gebbran2023_three_block_ADMM,Dimoulias2025_DroopAidedSE,Russell2025_DesignofVVC,Zhang2024_OptimizationofVVCdroopsetpoints} model the LV network by its positive-sequence, also known as ``single-phase equivalent''.
While it greatly reduces the size of the problem, especially in multi-period applications \cite{Gebbran2023_three_block_ADMM}, this simplification is not suitable for LV networks where the voltage unbalance is non-negligible, especially under high PV uptake which further exacerbates unbalance.
The methods in \cite{Inaolaji2021_LinDist3Flow_VVWO,Inaolaji2023_DOPFwithVVWOLinDist3Flow,Soltani2023_VVC,Quiertant2023_STATCOMSvsVVC} adopt another common simplification called Kron reduction, which reduces the $4 \times 4$ impedance matrix of self- and mutual impedances to an equivalent $3 \times 3$ matrix, assuming the neutral conductor is perfectly grounded, i.e., has zero grounding impedance.\footnote{Kron reduction is also valid if the neutral-to-ground voltage is uniform across all nodes in the network.}
However, when the neutral wire is not pervasively grounded or has a non-negligible grounding impedance, a Kron-reduced model may significantly underestimate voltage drops \cite{Urquhart2016_AccuracyofLVDN}.

In summary, as shown in Table~\ref{tab_summary_modellingstrategies}, none of the existing methods are suitable for VVWO in four-wire unbalanced LV DNs, which pose unique modeling challenges due to mutual coupling between phases, significant neutral currents, and large voltage unbalance, in addition to the myriad challenges introduced by the nonsmooth nature of the VVWC functions.
In fact, to the best knowledge of the authors, the formulation in \cite{Usman2020_SDPFourWireDPF} is the only known convex relaxation for the OPF in four-wire unbalanced LV DNs.
Nevertheless, because this formulation is based on semi-definite programming (SDP), whose OPF solution may not be exact, potential extensions to incorporate VVWO will likely be intractable since mixed-integer SDP is still a nascent field of research.

\begin{table}[t]
	\centering
	\caption{Summary of PF modeling strategies for VVWO in DNs.}
	\label{tab_summary_modellingstrategies}
	\renewcommand{\arraystretch}{1.25}
	\small

	\begin{tabular}{>{\raggedright\arraybackslash}m{2.2cm}
			>{\centering\arraybackslash}m{1.1cm}
			>{\centering\arraybackslash}m{1.1cm}
			m{2.7cm}}
		\toprule
		\textbf{Method} & \textbf{Accuracy} & \textbf{Efficiency} & \textbf{Papers} \\
		\midrule

		\textbf{Simplifications}
		& \xmark
		& \cmark$^{\dagger}$
		& Positive-sequence:  \cite{Olowu2021_VVC,Savasci2021_SOCP_VVC,Gebbran2023_three_block_ADMM,Dimoulias2025_DroopAidedSE,Russell2025_DesignofVVC,Ingebrigtsen2026_VVCinBESS,Zhang2024_OptimizationofVVCdroopsetpoints,Vespucci2026_PDIPMforVVC};
		Kron reduction: \cite{Inaolaji2021_LinDist3Flow_VVWO,Inaolaji2023_DOPFwithVVWOLinDist3Flow,Soltani2023_VVC,Quiertant2023_STATCOMSvsVVC} \\

		\midrule

		\textbf{Approximations}
		& \xmark
		& \cmark
		& LinDist3Flow: \cite{Inaolaji2021_LinDist3Flow_VVWO,Inaolaji2023_DOPFwithVVWOLinDist3Flow};
		Taylor-series models: \cite{Soltani2023_VVC,Russell2025_DesignofVVC}, no VVC deadband \cite{Russell2025_DesignofVVC}\\

		\midrule

		\textbf{Convex Relaxations}
		& \cmark$^{\ddagger}$
		& \xmark
		& SOCR/DistFlow: \cite{Savasci2021_SOCP_VVC} \\

		\midrule

		\textbf{Proposed Methods}
		& \cmark
		& \cmark
		& Full four-wire OPF (this work) \\

		\bottomrule
	\end{tabular}

	\begin{flushleft}
		{\footnotesize $\dagger$ Computationally efficient only when paired with linearizations.\\}
		{\footnotesize $\ddagger$ Accuracy only guaranteed under restrictive network conditions rarely satisfied in practice.}
	\end{flushleft}

\end{table}

\subsection{Paper Scope and Contributions}\label{subsec_introduction}

Against this background, this paper tackles the aforementioned challenges by introducing three high-accuracy and reliable optimization-based methods for VVWO in four-wire unbalanced networks that incorporate the nonsmooth VVWC functions without binary or integer variables.
The paper also identifies the algorithmic implementation framework that provides a good compromise between reliability and computational efficiency on real four-wire unbalanced LV networks.

The first method encodes these nonsmooth functions directly as user-defined functions using control flow, a feat enabled by many state-of-the-art algebraic modeling languages.
The second and third introduce bespoke \textit{smooth} approximations with tunable approximation errors to address potential numerical issues from nondifferentiability, improving reliability while maintaining accuracy and computational efficiency.
All three methods are evaluated on real four-wire unbalanced LV networks with varying rooftop solar uptake levels, including a 539-bus system with 302 smart inverters.
Results show that optimization-based non-incremental methods can outperform incremental (quasi-steady-state) approaches, and this work is the first to demonstrate accurate, reliable, and tractable VVWO on real four-wire unbalanced LV DNs.

This paper advances the state-of-the-art in the following ways:
\begin{itemize}
	\item It introduces three high-accuracy and reliable nonlinear VVWO formulations, including two tunable smooth approximations that provide explicit control over the balance between reliability, computational efficiency, and accuracy.
	\item It identifies, through rigorous numerical evaluation, an effective implementation strategy encompassing parameter tuning, linear solver selection, and numerical stabilization tools, and shows that the proposed exact nonlinear formulations vastly outperform the exact MINLP model in both reliability and computational efficiency.
	\item It demonstrates, for the first time, that optimization-based non-incremental formulations can exhibit greater algorithmic reliability than state-of-the-art incremental (quasi-steady-state) tools, particularly in large four-wire LV networks with high PV adoption.
\end{itemize}

The contributions of this paper can be classified into three categories: \textit{conceptual}, \textit{theoretical}, and \textit{implementation}.
On the conceptual front, this paper demonstrates that nondifferentiability can be tackled \textit{directly} with primal-dual interior-point methods (PDIPMs) in a \textit{reliable fashion}, without resorting to binary or integer variables.

On the theoretical front, it demonstrates that optimization-based non-incremental VVWC can be more reliable than the incremental control strategy, e.g., as implemented in OpenDSS \cite{OpenDSS}, commonly used in distribution system analysis.
In contrast, the non-incremental control architecture of the proposed nonlinear methods eliminates the outer iteration required in incremental control methods (see Fig.~\ref{fig:loops}) by embedding the VVWO directly into the network equations in a single unified system rather than through sequential updates.
This improves reliability in settings where strong coupling between phases and between active and reactive power can cause the iterative process in incremental methods to exhibit slow convergence or oscillatory behavior.

On the implementation front, this paper identifies the most suitable parameters, tools to improve numerical stability, and linear solvers that offer the best tradeoffs between reliability, accuracy, and computational efficiency.
The proposed methods can be instrumental in a wide variety of applications where VVWC is essential, such as DN hosting capacity, state estimation, dynamic operating envelopes \cite{Liu2024_RobustDOE}, multi-period DER coordination, customization of VVC and VWC inverter setpoints, and DN planning in general.

The paper is organized as follows. Section~\ref{sec_Formulation} details the mathematical modeling and the three different ways of incorporating the VVWC functions into the four-wire OPF, Section~\ref{sec_implementation} describes the case studies and the implementation setup, Section~\ref{sec_numericaleval} provides an exhaustive numerical evaluation of the proposed methods, and Section~\ref{sec_conclusion} concludes the paper. 
The present manuscript extends the authors' work published in \cite{Mhanna2026_VVWO_TPWRS} by providing additional derivations, numerical results, and sensitivity analyses.

\section{Mathematical modeling}\label{sec_Formulation}

As they are essential for interpreting and understanding the findings in this paper, the three keys concepts of \textit{reliability}, \textit{differentiability}, and \textit{smoothness}, are defined as follows.
\begin{definition}[Reliability] \label{def_reliability}
	Algorithmic reliability refers to the ability of an optimization or power flow algorithm to converge, in a numerically stable way, to a feasible solution across a broad range of operating conditions, nonlinearities, and initialization choices.
	Reliability is therefore evaluated by examining solver termination status, constraint satisfaction within prescribed tolerances, and consistency of convergence across multiple loading and voltage scenarios.
\end{definition}

\begin{definition}[Differentiability] \label{def_differentiability}
	A function $f: \mathcal{D} \to \mathbb{R}$ is differentiable on a domain $\mathcal{D} \subseteq \mathbb{R}^n$ if $\nabla f(x)$ exists for all $x \in \mathcal{D}$, and \textit{continuously} differentiable if $\nabla f(x)$ is a continuous function of $x$.
	Such a function also has the property
	\begin{align*}\label{eq_differentiability}
		\lim_{y \to 0} \frac{f(x + h) - f(x) - \nabla f(x)^{\mathsf T} y}{\lVert y \rVert} = 0, \quad \forall x \in \mathcal{D},
	\end{align*}
\end{definition}
where $\lVert \cdot \rVert$ is an arbitrary vector norm \cite{Bertsekas1999_NonlinearProgramming,Nocedal2006_NumericalOptimization}.

\begin{definition}[Smoothness] \label{def_smoothness}
	A function $f: \mathcal{D} \to \mathbb{R}$ is called \textit{smooth} on a domain $\mathcal{D} \subseteq \mathbb{R}^n$ if it is continuously differentiable on $\mathcal{D}$ \cite{Bertsekas1999_NonlinearProgramming,Nocedal2006_NumericalOptimization}.
	In general nonlinear programming, a problem is called smooth if the objective function and all constraint functions are continuously differentiable on the relevant domain.
\end{definition}

\subsection{OPF in four-wire unbalanced LV networks} \label{sec_FourWireOPF}

Consider an LV DN where the bus voltage vector $ \mathbf{U}_{i} $ stacks the phase voltages $ {U}_{ip}$
at bus \textit{i} such that $ p \in \mathcal{P}_{i}\subseteq \{ a,b,c,n \} $, and the branch current vector $ \mathbf{I}_{lij} $ stacks the conductor currents $ {I}_{lijp}$ flowing through phases $p$ of line \textit{l} from bus \textit{i} to bus \textit{j}.
The $\pi$-model of the branch is parameterized by the series impedance $\mathbf{Z}_{l}^{\rm s}$ and shunt admittance matrices
$\mathbf{Y}_{l}^{\rm fr} \text{ and } \mathbf{Y}_{l}^{\rm to}$.
All loads \textit{d} or generators \textit{g} are connected to buses in WYE configurations with the usual notations for powers and currents.
The reference bus is $i_{\rm ref}$, set up with a balanced phasor and perfect neutral grounding. The generator on the reference bus is $g_{\rm ref}$.
The symbol $\circ$ indicates the element-wise product and superscript $*$ indicates the conjugate.
Shunts $s$ such as capacitor banks and neutral grounding points are represented using admittance matrices $\mathbf{Y}_s$.
To avoid the nondifferentiability of the square root function, the magnitude of the phase-to-neutral voltage ${\mathbf{U}_{i}^{\rm m}}$ is expressed as
\begin{equation}
	({ \mathbf{U}_{i}^{\rm m}})^{2} = \Re( \mathbf{U}_{i} - {U}_{i,n})^{2} + \Im( \mathbf{U}_{i} - {U}_{i,n})^{2} ,  \label{voltage_magnitude_squared}
\end{equation}
along with
\begin{equation}
	\mathbf{U}_{i}^{\rm m}  \geq  \mathbf{0},     \label{voltage_magnitude_implicit_square_root}
\end{equation}
which restricts all magnitudes to be nonnegative.
To facilitate a direct comparison with existing PF solvers, this paper adopts a four-wire OPF problem that minimizes power import from the MV grid, which is equivalent to maximizing PV generation when it is available.
Consistent with \cite{Claeys2022_Fourwire}, this problem is stated as
\begin{subequations} \label{VVWO}
	\begin{align}
		{\mbox{minimize}} \ {P}^{\rm grid}  & & \label{objective} \\
		\text{subject to \ (\ref{voltage_magnitude_squared}), (\ref{voltage_magnitude_implicit_square_root}),} & & \label{voltage_magnitude} \\
		\mathbf{U}_{i} - \mathbf{U}_{j} & = \mathbf{Z}_{l}^{\rm s} \mathbf{I}_{lij}^{s}   \label{ohms_law}, \\
		\mathbf{I}_{lij} & = \mathbf{Y}_{l}^{\rm fr}\mathbf{U}_{i} + \mathbf{I}_{lij}^{s}  , \label{current_from}\\
		\mathbf{I}_{lji} & = \mathbf{Y}_{l}^{\rm to}\mathbf{U}_{j} - \mathbf{I}_{lij}^{s} ,  \label{current_to}\\
		\mathbf{S}_{d} & =  (\mathbf{U}_{i}- {U}_{i,n})\circ \mathbf{I}_{d}^{\ast}  , \label{load_power}\\
		\mathbf{S}_{g} & = \mathbf{P}_{g} + j \mathbf{Q}_{g} =   (\mathbf{U}_{i}- {U}_{i,n}) \circ \mathbf{I}_{g}^{\ast} , \label{gen_power}\\
		\mathbf{1}^{\text{T}} \mathbf{I}_{d}& = 0  , \label{load_current_conservation}\\
		\mathbf{1}^{\text{T}}  \mathbf{I}_{g}& = 0  , \label{gen_current_conservation}\\
		P^{\text{grid}} & =\mathbf{1}^\text{T}\mathbf{P}_{g_{\rm ref}}, \ \label{eq_p_grid} \\
		\mathbf{U}_{i_{\rm ref}} &= \begin{bmatrix}
			1 & e^{-2\pi/3} & e^{2\pi/3} & 0
		\end{bmatrix}^\text{T} ,\\
		\sum_{lij}  \mathbf{I}_{lij} + \sum_{di} \mathbf{I}_{d} & -  \sum_{gi}  \mathbf{I}_{g} +\sum_{si} \mathbf{Y}_s \mathbf{U}_i  = 0 , \label{bus_kcl} \\
		\mathbf{P}_{g}^{\rm min} & \leq  \mathbf{P}_{g} \leq  \mathbf{P}_{g}^{\rm max}  \label{pg_bounds},\\
		\mathbf{Q}_{g}^{\rm min} & \leq  \mathbf{Q}_{g} \leq  \mathbf{Q}_{g}^{\rm max}  \label{qg_bounds},\\
		\mathbf{S}_{g} \circ \mathbf{S}_{g}^*  & \leq  \mathbf{S}_{g}^{\rm max} \circ \mathbf{S}_{g}^{\rm max}  \label{thermal_limit},\\
		\mathbf{U}_{i}^{\rm min} \circ   \mathbf{U}_{i}^{\rm min} &\leq
		\mathbf{U}_{i}^{\rm m} \circ (\mathbf{U}_{i}^{\rm m})^*  \leq
		\mathbf{U}_{i}^{\rm max} \circ   \mathbf{U}_{i}^{\rm max}, \label{voltage_bounds}
	\end{align}
\end{subequations}
where $\mathbf{I}_{lij}^{s} $ is the series current vector of line \textit{l}  from bus \textit{i} to bus \textit{j}.

\subsection{VVWC as Nonsmooth NLP Problem (\nonsmooth )} \label{subsec_nonsmooth}

\begin{figure}[tbh]
	\begin{minipage}{0.45\columnwidth}

\begin{tikzpicture}[x=0.5pt,y=0.5pt,yscale=-1,xscale=1]

\draw  (23,64.83) -- (211.5,64.83)(39.02,13) -- (39.02,118) (204.5,59.83) -- (211.5,64.83) -- (204.5,69.83) (34.02,20) -- (39.02,13) -- (44.02,20)  ;
\draw[line width=0.75pt]    (61.8,31) -- (101,64.7) ;

\draw[line width=0.75pt]    (130,65) -- (168.5,98) ;

\draw[line width=0.75pt]    (101,65) -- (130,65) ;

\draw[line width=0.75pt]    (62,31) -- (39,31) ;

\draw[line width=0.75pt]    (190,98) -- (168.3,98) ;

\draw[thin]  [dash pattern={on 0.84pt off 2.51pt}]  (39.5,98) -- (168.5,98) ;

\draw (65,62) -- ++ (0,5);
\draw (101,62) -- ++ (0,5);
\draw (130,62) -- ++ (0,5);
\draw (168.5,62) -- ++ (0,5);

\draw (33,59) node [scale=0.8]  {$0$};
\draw (65,79) node [scale=0.8]  {$U_{1}$};
\draw (101,79) node [scale=0.8]  {$U_{2}$};
\draw (130,79) node [scale=0.8]  {$U_{3}$};
\draw (168.5,79) node [scale=0.8]  {$U_{4}$};
\draw (33,96) node [scale=0.8]  {$\underline{q}$};
\draw (33,29) node [scale=0.8]  {$\overline{q}$};
\draw (222,78) node [scale=0.8]  {$\text{ V}$};
\draw (36,0) node [scale=0.8]  {$\text{ VAr}$};

\end{tikzpicture}
		\\{\small{{a) Generic VVC function.}} }

	\end{minipage}
	\begin{minipage}{0.45\columnwidth}

\begin{tikzpicture}[x=0.5pt,y=0.5pt,yscale=-1,xscale=1]

\draw  (23,72) -- (211.5,72)
(39.02,13) -- (39.02,118) 
(204.5,67) -- (211.5,72) -- (204.5,77) (34.02,20) -- (39.02,13) -- (44.02,20)  ;

\draw[line width=0.75pt]    (101,31) -- (130,64.7) ;


\draw[line width=0.75pt]    (130,65) -- (168.3,65) ;

\draw[line width=0.75pt]    (39,31) -- (101,31) ;


\draw[thin]  [dash pattern={on 0.84pt off 2.51pt}]  (39.5,64.83) -- (140,64.83) ;

\draw (101,70) -- ++ (0,5);
\draw (130,70) -- ++ (0,5);

\draw (34,82) node [scale=0.8]  {$0$};
\draw (101,85) node [scale=0.8]  {$U_{5}$};
\draw (130,85) node [scale=0.8]  {$U_{6}$};
\draw (33,64.83) node [scale=0.8]  {$\underline{p}$};
\draw (33,29) node [scale=0.8]  {$\overline{p}$};
\draw (222,85) node [scale=0.8]  {$\text{ V}$};
\draw (36,0) node [scale=0.8]  {$\text{ W}$};

\end{tikzpicture}
		\\{\small{{b) Generic VWC function.}}}
	\end{minipage}
	\caption{ (a) VVC and (b) VWC settings.}\label{fig:VVWC_settings}
\end{figure}
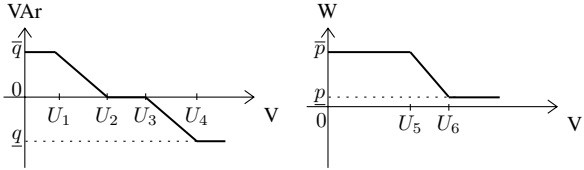

VVC helps regulate voltages by locally injecting or absorbing reactive power following a droop setting illustrated in Fig.~\ref{fig:VVWC_settings}a.
A generic VVC function can be expressed as
\begin{equation}
	Q_{gp}(U_{i}^{\rm m}) = \begin{cases}
			\overline{q} & \text{if} \quad  U_{ip}^{\rm m} \leq {U}_{1},\\
			{\overline{q}} \frac{( U_{2}- U_{ip}^{\rm m} )}{( U_{2}-U_{1} )} & \text{if} \quad {U}_{1} <  U_{ip}^{\rm m} < {U}_{2},\\
			0 & \text{if} \quad {U}_{2}\leq  U_{ip}^{\rm m}\leq {U}_{3},\\
			\underline{q} \frac{(  U_{ip}^{\rm m}-U_{3} )}{( U_{4}-U_{3} )} & \text{if} \quad {U}_{3} <  U_{ip}^{\rm m} < {U}_{4},\\
			\underline{q} & \text{if} \quad {U}_{4} \leq U_{ip}^{\rm m},
	\end{cases}
	\label{VVC_constraint}
\end{equation}
where $Q_{gp}$ is the reactive power output exchanged on phase \textit{p}, $\overline{q}$ is the maximum reactive power output (normally defined as a percentage of the inverter's maximum capacity $S_{g}^{\rm max}$), and $U_{ip}^{\rm m}$ is the phase-to-neutral voltage magnitude measured at bus \textit{i} and phase $p$.

Similarly, VWC helps regulate voltages by locally reducing active power dispatch from PV inverters, $P_{gp}$, following a droop setting shown in Fig.~\ref{fig:VVWC_settings}b.
A generic VWC function can be expressed as
\begin{equation}
	{P_{gp}}(U_{i}^{\rm m}) \leq \begin{cases}
		\overline{p} & \text{if} \quad  U_{ip}^{\rm m} \leq {U}_{5},\\
		\frac{{(\underline{p}}- \overline{p})(U_{ip}^{\rm m} -U_{5} )}{U_{6}-U_{5}} + \overline{p} & \text{if} \quad {U}_{5} <  U_{ip}^{\rm m} < {U}_{6},\\
		\underline{p} & \text{if} \quad {U}_{6} \leq  U_{ip}^{\rm m},
	\end{cases}\\
	\label{VWC_constraint}
\end{equation}
where the inequality is crucial to avoid trivial infeasibility in cases where both VVC and VWC are active and the power generated by the PV panels is at the nominal rating of the inverter.
For example if a 5 kVA inverter operates at its rated power of 5 kW, and if the voltage magnitude at the connection point is such that reactive power has to be exchanged according the VVC settings ($U_{ip}^{\rm m}$ has a value below $ U_{2}$ or above $ U_{3}$), the inverter's kVA limit \eqref{thermal_limit} cannot be satisfied.
Since a solar inverter can always reduce its output by deviating from the maximum power point, the inequality in \eqref{VWC_constraint} is necessary to ensure feasible operation while maintaining voltage regulation.
It should be noted that the VVC and VWC are not modeled as bus constraints but as constraints on a generator connected to a bus.

The two functions in \eqref{VVC_constraint} and \eqref{VWC_constraint} are not differentiable at their breakpoints and are not continuously differentiable, as their gradients are not continuous over their respective domains.
Consequently, smooth approximations are introduced next to ensure compatibility with assumptions made by gradient-based nonlinear optimization solvers.

\subsection{VVWC as a Smooth NLP Problem} \label{subsec_smooth}

In general, a PWL function can be approximated by a smooth counterpart using either a single closed-form function or through a collection of smooth curves called splines.
The aim is to ensure that the function is continuously differentiable across its whole domain.
Therefore for the closed-form option, the choice of function is usually one that is \textit{analytic}.
One such function is the \textit{softplus}, which is commonly used in machine learning applications to approximate the rectified linear unit, i.e., ``ReLU'' function.

On the other hand, a \textit{spline} is a sequence of several smaller curve segments that join smoothly.
Each segment has its own polynomial, usually of low order, and all segments are connected so that the whole spline is differentiable.
The lowest order spline that satisfies the smoothness and differentiability requirements is the \textit{quadratic} Bézier spline.

\subsubsection{Softplus (\softplus)} \label{subsec_softplus}

This section details how to approximate the VVC and VWC functions in \eqref{VVC_constraint} and \eqref{VWC_constraint} using a single smooth function.
Consider the ReLU function \(f(x) = \max(0,x)\) with \(\dom f = \mathbb{R} \).\footnote{Function \(f(x)\) is not differentiable at \(x = 0 \).}
For any positive $\epsilon$, \(f(x)\) can be approximated by the smooth \emph{softplus} function $f^{\epsilon}(x) = \epsilon \log(1+\exp(x/\epsilon) ) $, with approximation error bounded by $f^{\epsilon}(x) - \epsilon\log(2) \leq f(x) \leq f^{\epsilon}(x)  $ \cite{Chen_smooth_approximation}.
The approximation error therefore decreases as $\epsilon$ decreases.
In general, any PWL function $f(x)$ can be written as
\[
f(x) = c + a_0 x + \sum_{i=1}^{n} (a_i - a_{i-1})\,\max(x - t_i , 0),
\]
where $t_1 < t_2 < \ldots < t_n$ are the breakpoints and $a_0, a_1, \cdots, a_n$ are the slopes of $f(x)$ in each interval.
The VVC function in \eqref{VVC_constraint} can be rewritten as
\begin{align*}
	Q_{gp}(U_{i}^{\rm m}) =  \overline{q} & + (a_1 - a_0)\max(U_{i}^{\rm m}-U_1,0) \\
	& + (a_2 - a_1)\max(U_{i}^{\rm m}-U_2,0) \\
	& + (a_3 - a_2)\max(U_{i}^{\rm m}-U_3,0) \\
	& + (a_4 - a_3)\max(U_{i}^{\rm m}-U_4,0),
\end{align*}
where $a_0 = 0$, $a_1 = -\frac{\overline{q}}{U_2-U_1}$, $a_2 = 0$, $a_3 = \frac{\underline{q}}{U_4-U_3}$, and $a_4 = 0$.
Functions \eqref{VVC_constraint} and \eqref{VWC_constraint} can now each be approximated with softplus counterparts
\begin{align}
	Q_{gp}^{\rm soft}&(U_{i}^{\rm m}) =  \overline{q} \nonumber \\
	& + (a_1 - a_0)\epsilon_{\rm vvc} \log\left(1+\exp\left(\frac{U_{i}^{\rm m}-U_{1}}{\epsilon_{\rm vvc}}\right)\right) \nonumber \\
	& + (a_2 - a_1)\epsilon_{\rm vvc} \log\left(1+\exp\left(\frac{U_{i}^{\rm m}-U_{2}}{\epsilon_{\rm vvc}}\right)\right) \nonumber \\
	& + (a_3 - a_2)\epsilon_{\rm vvc} \log\left(1+\exp\left(\frac{U_{i}^{\rm m}-U_{3}}{\epsilon_{\rm vvc}}\right)\right) \nonumber \\
	& + (a_4 - a_3)\epsilon_{\rm vvc} \log\left(1+\exp\left(\frac{U_{i}^{\rm m}-U_{4}}{\epsilon_{\rm vvc}}\right)\right), \label{softplus_VVC}
\end{align}
and
\begin{align}
	P_{gp}^{\rm soft}&(U_{i}^{\rm m}) \leq \overline{p} \nonumber \\
	& + (b_1 - b_0)\epsilon_{\rm vwc} \log\left(1+\exp\left(\frac{U_{i}^{\rm m}-U_{5}}{\epsilon_{\rm vwc}}\right)\right) \nonumber \\
	& + (b_2 - b_1)\epsilon_{\rm vwc} \log\left(1+\exp\left(\frac{U_{i}^{\rm m}-U_{6}}{\epsilon_{\rm vwc}}\right)\right), \label{softplus_VWC}
\end{align}
where $b_0 = 0$, $b_1 = \frac{\underline{p}-\overline{p}}{U_{6} - U_{5}}$, $b_2 = 0$, and $\epsilon_{\rm vvc} = \epsilon_{\rm vwc} = \epsilon/V_{\rm base}$.

The smooth VVC and VWC formulations in \eqref{softplus_VVC}–\eqref{softplus_VWC} follow from expressing each PWL droop segment as a linear combination of shifted ReLU functions and replacing each ReLU with its softplus approximation to produce a single, (twice) continuously differentiable smooth function with slopes that exactly match the original droop curves.

\subsubsection{Spline (\spline)} \label{subsec_spline}
This section details how to approximate the VVC and VWC functions in \eqref{VVC_constraint} and \eqref{VWC_constraint} with a quadratic Bézier spline made of several quadratic Bézier curves.
A quadratic Bézier curve is a single curve segment defined by three control points $P_0$, $P_1$, and $P_2$, and its equation is
\begin{equation*}
	B(t) = (1 - t)^2 P_0 + 2(1 - t)t P_1 + t^2 P_2,
	\qquad t \in [0, 1].
\end{equation*}
To approximate a breakpoint $P^{k}_0=(U_{k}, Q_{gp}(U_{k}))$ of the VVC function by a quadratic Bézier curve, two equidistant control points $P^{k}_1=(U_{k}-\Delta U, Q_{gp}(U_{k}-\Delta U))$ and $P^{k}_2=(U_{k}+\Delta U, Q_{gp}(U_{k}+\Delta U))$ are introduced.
Because points $P^{k}_1$ and $P^{k}_2$ are of equal distance to $P^{k}_0$, the equation of the quadratic Bézier can be written entirely as a function of the variables $U_{i}^{\rm m}$ as
\begin{align*}
	Q_{gp}^{\rm spl}(U_{i}^{\rm m}) = & (1 - \beta)^2 Q_{gp}(U_{k}) + 2(1 - \beta)\beta Q_{gp}(U_{k}-\Delta U)
\end{align*}
\begin{align} \label{eq_Qspline}
	+ & \beta^2 Q_{gp}(U_{k}+\Delta U),
\end{align}
where
\begin{equation*}
	\beta = \frac{U_{k} + \Delta U - U_{i}^{\rm m}}{2 \Delta U}.
\end{equation*}

The breakpoints of the VWC function in \eqref{VWC_constraint} can be similarly approximated as
\begin{align*}
	P_{gp}^{\rm spl}(U_{i}^{\rm m}) \leq & (1 - \beta)^2 P_{gp}(U_{k}) + 2(1 - \beta)\beta P_{gp}(U_{k}-\Delta U)
\end{align*}
\begin{align} \label{eq_Pspline}
	+ & \beta^2 P_{gp}(U_{k}+\Delta U).
\end{align}
The approximation error in $Q_{gp}^{\rm spl}(U_{i}^{\rm m})$ and $P_{gp}^{\rm spl}(U_{i}^{\rm m})$ can then be controlled by tuning parameter $\Delta U$.
For consistency and fairness in comparison, the approximation error of the splines in \eqref{eq_Qspline} and \eqref{eq_Pspline} can be related to $\epsilon_{\rm vvc}$ and $\epsilon_{\rm vwc}$ of the softplus function through
\begin{align}
	\Delta U^{\rm vvc} = \frac{4\epsilon_{\rm vvc} (U_{2} - U_{1})}{\overline{q}},
\end{align}
and
\begin{align}
	\Delta U^{\rm vwc} = \frac{4\epsilon_{\rm vwc} (U_{6} - U_{5})}{\underline{p} - \overline{p}},
\end{align}
respectively.

\begin{figure}[t!]
	\centering
	\begin{subfigure}{1\columnwidth}
		\includegraphics[width=1\columnwidth]{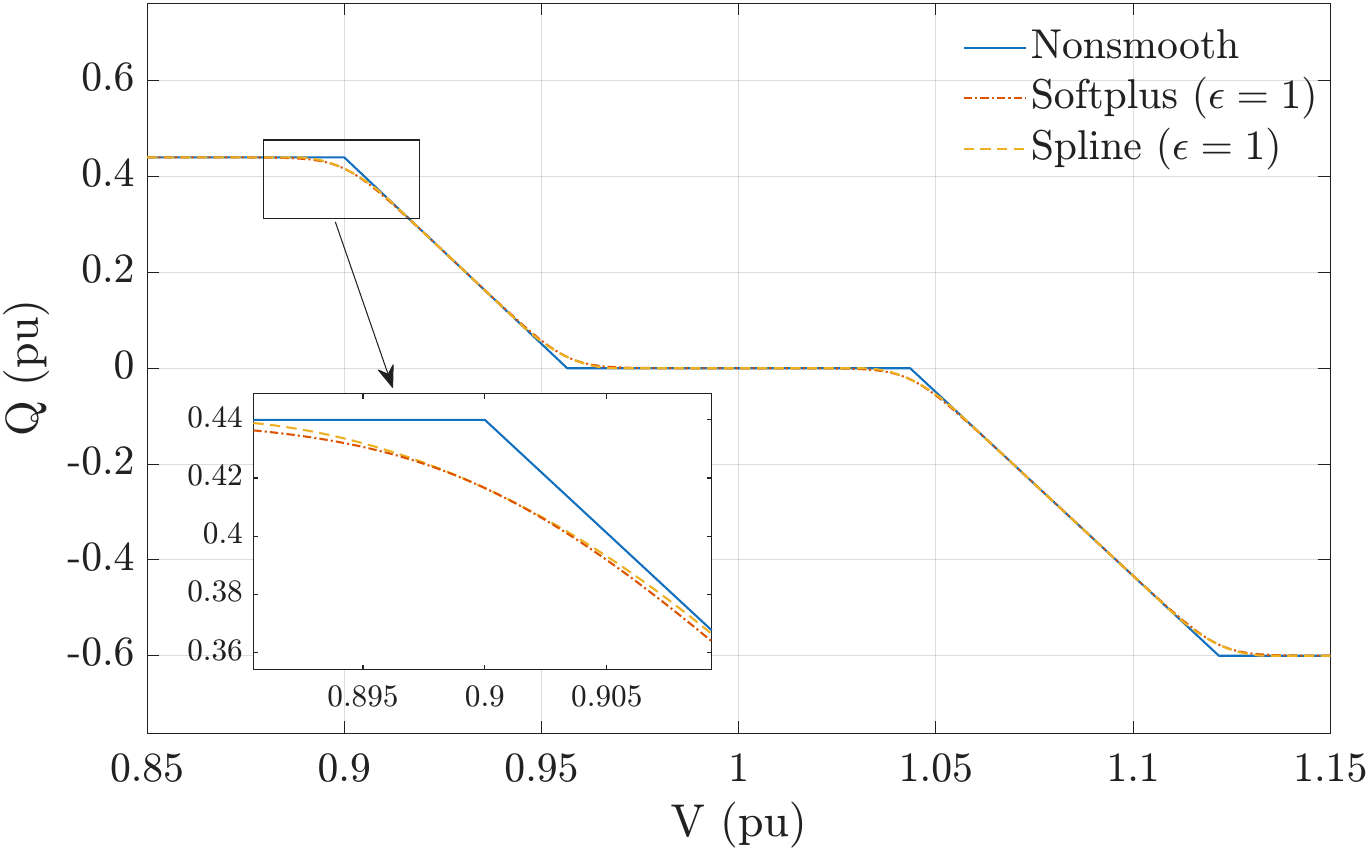}
		\caption{VVC}
		\label{fig_VVCmodels}
	\end{subfigure}
	\hfill
	\begin{subfigure}{1\columnwidth}
		\includegraphics[width=1\columnwidth]{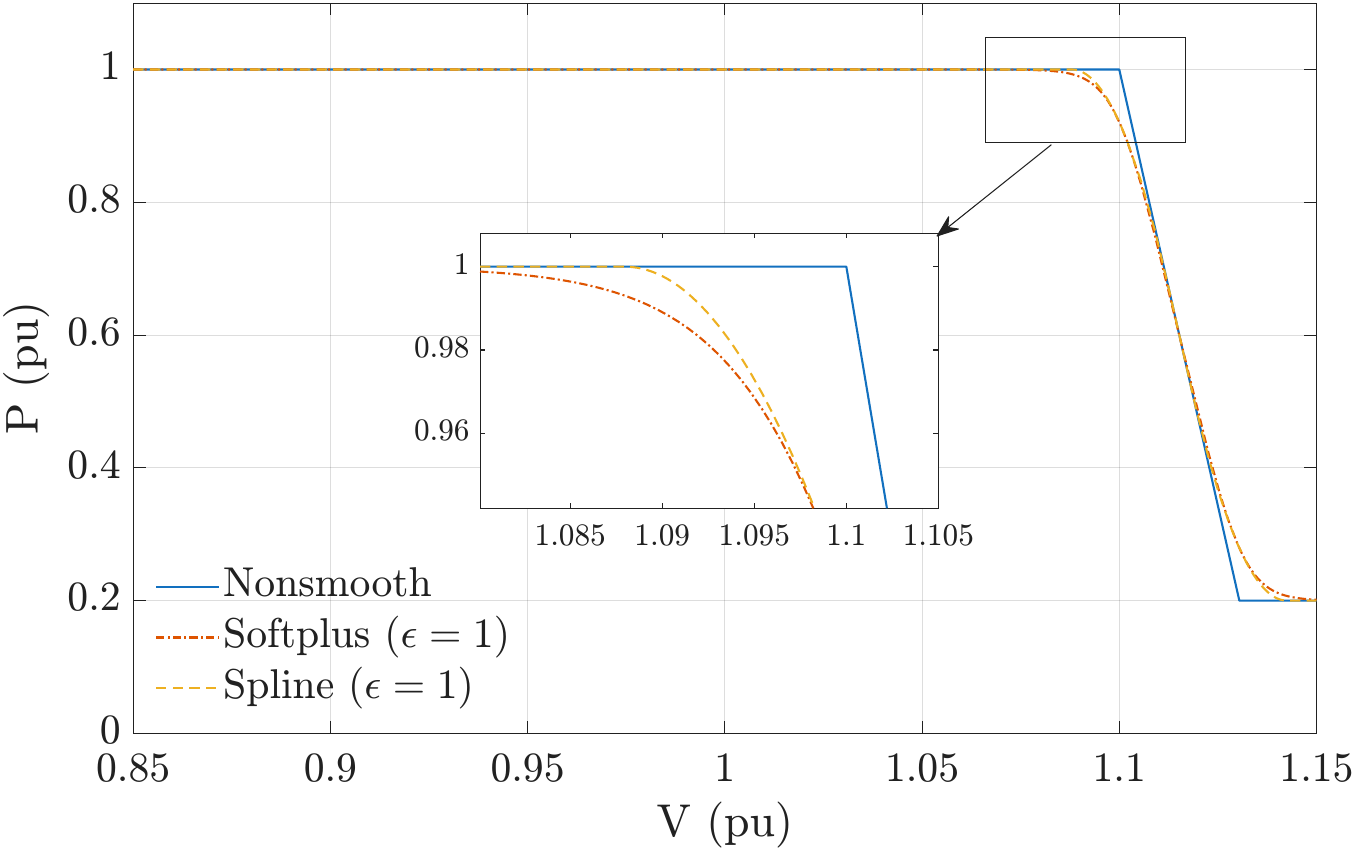}
		\caption{VWC}
		\label{fig_VWCmodels}
	\end{subfigure}
	\caption{Illustration of the softplus and spline approximations for $\epsilon = 1$.}
	\label{fig_VWWCmodels}
\end{figure}

Fig. \ref{fig_VVCmodels} depicts the softplus \eqref{softplus_VVC} and spline \eqref{eq_Qspline} VVC approximations and how they compare to the nonsmooth formulation in \eqref{VVC_constraint} for $\epsilon = 1$.
Fig. \ref{fig_VWCmodels} depicts the softplus \eqref{softplus_VWC} and spline \eqref{eq_Pspline} VWC approximations and how they compare to the nonsmooth formulation in \eqref{VWC_constraint} for $\epsilon = 1$.

\subsection{VVWO as an MINLP Problem (\minlp )} \label{subsec_minlp}

The aim of this approach is to encode \eqref{VVC_constraint} and \eqref{VWC_constraint} with binary or integer variables, making the problem amenable to being solved to \textit{exactness} using an NLBB method.
However, instead of assigning one binary variable to each PWL segment in \eqref{VVC_constraint} and \eqref{VWC_constraint}, as is the case in the \textit{incremental}\footnote{Not to be confused with incremental control.} or \textit{convex combination} formulations, the approach in this paper uses a \emph{logarithmic} formulation \cite{huchette2019nonconvex} which encodes \eqref{VVC_constraint} with only 3 binary variables instead of 5, and \eqref{VWC_constraint} with 2 binary variables instead of 3, thereby improving computational efficiency.

In more detail, let $\{(U_k, q_k)\}_{k=1}^{N}$ denote the breakpoints and corresponding reactive power values obtained from the VVC function in \eqref{VVC_constraint}.
For each bus $i$, inverter $g$, and phase $p$, variables $\lambda_{gp,k} \ge 0$
and binary variables $z_{gp,j} \in \{0,1\}$ for
$j = 1,\dots,\lceil \log_2(N-1) \rceil$ are introduced.
The VVC constraint can then be written as
\begin{subequations}
	\label{eq:vvc_system}
	\begin{align}
		U_{ip}^{\rm m} &= \sum_{k=1}^{N} U_k \lambda_{gp,k}, \label{eq:log_vvc_u} \\
		Q_{gp} &= \sum_{k=1}^{N} q_k \lambda_{gp,k}, \label{eq:log_vvc_q} \\
		\sum_{k=1}^{N} \lambda_{gp,k} &= 1,\qquad \lambda_{gp,k} \ge 0, \label{eq:log_vvc_sum} \\
		\intertext{together with the logarithmic constraints}
		\sum_{k \in [N] \setminus L^j} \lambda_{gp,k} &\le z_{gp,j}, \qquad \forall j, \label{eq:log_vvc_z1} \\
		\sum_{k \in [N] \setminus R^j} \lambda_{gp,k} &\le 1 - z_{gp,j}, \qquad \forall j, \label{eq:log_vvc_z2} \\
		z_{gp,j} &\in \{0,1\}, \label{eq:log_vvc_z3}
	\end{align}
\end{subequations}
The sets $L^j$ and $R^j$ depend on the Gray code associated with the logarithmic formulation \cite{huchette2019nonconvex} and ensure that the combination variables $\lambda_{gp,k}$ interpolate along a single PWL segment.
This encoding is exact and produces a compact MINLP representation of the VVC constraint.

Similarly, let $\{(U_k, p_k)\}_{k=1}^{M}$ denote the voltage breakpoints and corresponding active power values of the VWC droop in \eqref{VWC_constraint}.
By introducing continuous variables $\lambda^{\mathrm{W}}_{gp,k} \ge 0$ with $\sum_{k=1}^{M} \lambda^{\mathrm{W}}_{gp,k} = 1$ and binary variables $z^{\mathrm{W}}_{gp,j} \in \{0,1\}$ for $j = 1,\dots,\lceil \log_2(M-1) \rceil$, the VWC constraint is encoded as
\begin{subequations}
	\label{eq:vwc_logarithmic}
	\begin{align}
		U_{ip}^{\rm m} &= \sum_{k=1}^{M} U_k \lambda^{\mathrm{W}}_{gp,k}, \label{eq:vwc_u} \\
		P_{gp} &\le \sum_{k=1}^{M} p_k \lambda^{\mathrm{W}}_{gp,k}, \label{eq:vwc_p} \\
		\sum_{k=1}^{M} \lambda^{\mathrm{W}}_{gp,k} &= 1,\qquad \lambda^{\mathrm{W}}_{gp,k} \ge 0, \label{eq:vwc_sum} \\
		\intertext{along with the logarithmic constraints}
		\sum_{k \in [M] \setminus L^{j,\mathrm{W}}} \lambda^{\mathrm{W}}_{gp,k} &\le z^{\mathrm{W}}_{gp,j}, \qquad \forall j, \label{eq:vwc_z1} \\
		\sum_{k \in [M] \setminus R^{j,\mathrm{W}}} \lambda^{\mathrm{W}}_{gp,k} &\le 1 - z^{\mathrm{W}}_{gp,j}, \qquad \forall j, \label{eq:vwc_z2} \\
		z^{\mathrm{W}}_{gp,j} &\in \{0,1\}. \label{eq:vwc_z3}
	\end{align}
\end{subequations}
This formulation preserves the necessary inequality structure of the VWC constraint, which avoids trivial infeasibility when VVC and VWC are active simultaneously.
As in the VVC case, the logarithmic formulation yields an exact representation that requires only $\lceil \log_2(M-1) \rceil$ binary variables
per inverter and phase.

Since nonlinear branch-and-bound (NLBB) solves a nonlinear relaxation at every node, the total number of binary variables is a primary driver of computational cost.
It is therefore useful to introduce a simple complexity indicator \(\mathcal{O}\) to quantify the computational effort associated with each VVWO formulation, by defining
\[
\mathcal{O} := N_{\mathrm{bin}}^{\mathrm{VVWO}},
\]
i.e., the total number of binary variables introduced by the VVWO encoding.
For example, an incremental formulation of the VVC and VWC curves requires one
binary variable per PWL segment, which yields
\[
\mathcal{O}_{\mathrm{inc}}
= 5\, n_{\mathrm{inv}}^{\mathrm{VVC}}
+ 3\, n_{\mathrm{inv}}^{\mathrm{VWC}}.
\]
In contrast, the logarithmic formulation used in this work requires only $\left\lceil \log_{2}(6-1) \right\rceil = 3$ binary variables for VVC and $\left\lceil \log_{2}(4-1) \right\rceil = 2$ binary variables for VWC per inverter and phase. This reduction decreases \(\mathcal{O}\) and produces a much smaller NLBB branching tree.

Although this reduction directly improves computational tractability in NLBB MINLP solvers, this benefit is not universal.
In MILP applications that use branch-and-cut algorithms, performance is often problem dependent and a smaller number of binary variables does not necessarily lead to faster solution times \cite{huchette2019nonconvex}.

\subsection{Summary of Problem Specifications}
Table \ref{tab:comparison} summarizes the features and feasible sets for the proposed four VVWO problems described above.

\begin{table*}[tbh]
	\centering
	\caption{Summary of the VVWO Problem Specifications, Mathematical Properties, and Solvers.}
	\label{tab:comparison}
	\begin{tabular}{l l l l l}
		\cmidrule[0.08em](l){2-5}
		& \textbf{NLP nonsmooth}
		& \textbf{NLP softplus}
		& \textbf{NLP (quadratic) spline}
		& \textbf{MINLP} \\
		{}
		& \nonsmooth{}
		& \softplus{}
		& \spline{}
		& \minlp{} \\
		\midrule
		Smoothness
		& no
		& yes
		& yes
		& yes (when relaxed to continuous) \\
		Exactness
		& yes
		& yes (depending on $\epsilon$)
		& yes (depending on $\epsilon$)
		& yes \\
		Solver
		& \textsc{Ipopt} \cite{ipopt}
		& \textsc{Ipopt} \cite{ipopt}
		& \textsc{Ipopt} \cite{ipopt}
		& \textsc{Juniper} \cite{juniper} \\
		Smoothing setting
		& N/A (Sec.~\ref{subsec_nonsmooth})
		& $\epsilon$ (Sec.~\ref{subsec_softplus})
		& $\epsilon$ (Sec.~\ref{subsec_spline})
		& N/A (Sec.~\ref{subsec_minlp})  \\
		\midrule
		\parbox[c]{3cm}{Four-wire OPF constraints\\
			\cite{Claeys2022_Fourwire,PMD}}
		& \eqref{voltage_magnitude}--\eqref{voltage_bounds}
		& \eqref{voltage_magnitude}--\eqref{voltage_bounds}
		& \eqref{voltage_magnitude}--\eqref{voltage_bounds}
		& \eqref{voltage_magnitude}--\eqref{voltage_bounds} \\
		\midrule
		VVWC constraints
		& \eqref{VVC_constraint}--\eqref{VWC_constraint}
		& \eqref{softplus_VVC}--\eqref{softplus_VWC}
		& \eqref{eq_Qspline}--\eqref{eq_Pspline}
		& logarithmic PWL \cite{huchette2019nonconvex} \\
		VVWC encoding
		& user-defined (control flow)
		& user-defined (analytic)
		& user-defined (control flow)
		& binary \\
		\bottomrule
	\end{tabular}

\end{table*}

\section{Input data and implementation}\label{sec_implementation}

\subsection{Input Data and Assumptions}

The impact of network size on convergence is evaluated on three networks: \smallnw{}, \mediumnw{}, and \largenw{}, with 28, 48, and 302 end-users, respectively.
These networks were carefully selected from a dataset of 128 four-wire LV DNs, representative of feeders in the UK \cite{Nando_LV_feeders} up-sampled to four-wire \cite{fourwiredataset}.
With four phases per bus, these networks are somewhat comparable in model size to positive-sequence models with 120, 396, and 2,156 buses, respectively.

Similarly, to assess the impact of the number of VVWC functions on reliability and computational performance, three PV adoption scenarios are considered for each network: Low DER, Medium DER, and High DER.
The Low PV adoption scenario assumes that 33\% of end-users are equipped with PV systems.
The Medium PV adoption scenario assumes that 66\% of its end-users are equipped with PV systems, and the High PV adoption scenario assumes that 100\% of its end-users are equipped with PV systems.
Only single-phase PV inverters are considered in this work, and they are allocated to the same phase a customer is connected to.
All inverters are assumed to have a rating of 5.25 kVA, and all end-users with PV systems are assumed to provide VVWC.

The evaluation is performed during daylight hours when PV systems are able to export power, and VVWC functions are actively mitigating overvoltages.
Specifically, demand and PV profiles with half-hourly resolution between 8AM and 8PM are obtained from smart-meter data provided by AusNet as part of Project EDGE \cite{EDGE2023}.
As a result, 25 snapshot optimization problems are solved for each network and each PV adoption scenario.

The values of the breakpoints in the VVC function in Fig. \ref{fig_VVCmodels} are $U_{1} = 207 \ \mathrm{V}$, $U_{2} = 220 \ \mathrm{V}$, $U_{3} = 240 \ \mathrm{V}$, $U_{4} = 258 \ \mathrm{V}$ with $V_{\rm base} = 230 \ \mathrm{V}$.
The associated reactive power response is $\overline{q} = 44\%$ and $\underline{q} = 60\%$ of inverter rated capacity, in accordance with Australian standard AS/NZS 4777.2 \cite{ASNZ}.
Similarly, the values of the breakpoints in the VWC function in Fig. \ref{fig_VWCmodels} are $ U_{5} = 253 \ \mathrm{V}$, $U_{6} = 260 \ \mathrm{V}$, and the active power setpoints are $\overline{p} = 100\%$ and $\underline{p} = 20\%$ of inverter rated capacity \cite{ASNZ}.

Table \ref{tab:feeder_summary} provides a summary of the different test cases and shows the highest voltage observed in the networks when no VVWC is active and the PVs are dispatching under a unity power factor (UPF), i.e., without any reactive power support, and without any active power curtailment.

\begin{table}[h]
	\centering
	\caption{Characteristics of the three considered LV networks under UPF for three levels of PV adoption. The reported voltage is phase-to-neutral.}
	\label{tab:feeder_summary}
	\begin{tabular}{llccc}
		\toprule
		Network & Scenario & Total PV capacity & Max. voltage & \# VVWC \\
		\midrule
		\multirow{3}{*}{\smallnw{}} & Low & 52.5\,kW & 1.072 pu & 10 \\
		& Medium &94.5\,kW & 1.079 pu & 18  \\
		& High & 147\,kW & 1.104 pu& 28 \\
		\hline
		\multirow{3}{*}{\mediumnw{}} & Low & 84\,kW & 1.050 pu & 16        \\
		& Medium &168\,kW &1.123 pu& 32     \\
		& High  & 252\,kW & 1.178 pu & 48    \\
		\hline
		\multirow{3}{*}{\largenw{}} & Low & 530.25\,kW & 1.190 pu & 101     \\
		& Medium & 1055.2\,kW & 1.266 pu & 201    \\
		& High  & 1585.5\,kW & 	1.419 pu & 302     \\
		\bottomrule
	\end{tabular}%
\end{table}

\subsection{Computational setup}

\textsc{Julia} v1.9.2 \cite{Bezanson2017_Julia} is used as a programming language along with \textsc{JuMP} (v.0.23.2) \cite{DunningHuchetteLubin2017_JuMP} as a mathematical modeling layer for the optimization models.
The four-wire current–voltage OPF formulation in \textsc{PowerModelsDistribution.jl} \cite{PMD} is adopted as the foundational mathematical model for the four-wire OPF.
The evaluations are conducted on a 64-bit Intel Core i9-13900K (5.8 GHz) with 32 threads and 64 GB of RAM (DDR5, 6000 MHz).

All NLP problems are solved using \textsc{Ipopt} \cite{ipopt} (v1.4.2) with linear solver \textsc{MA27} \cite{montoison-orban-hsl-2021} and Math Kernel Library \textsc{MKL}\footnote{ \url{https://github.com/JuliaLinearAlgebra/MKL.jl?tab=readme-ov-file}} for the underlying linear algebra.
MA27 was deemed, based on rigorous and pragmatic evaluation, to offer the best tradeoff between reliability and computational efficiency compared to other linear solvers such as MUMPS, MA57, MA77, MA86, and MA97.
The MINLP problems are solved using \textsc{Juniper} (v0.9.1) \cite{juniper} with \textsc{PiecewiseLinearOpt.jl}~\cite{huchette2019nonconvex} (v0.4.1) for encoding the logarithmic formulations in Section~\ref{subsec_minlp}.
The same PDIPM solver and linear solvers are used for solving each node in the NLBB tree.

\subsection{Implementation}

A seamless implementation of the proposed \nonsmooth{}, \softplus{}, and \spline{} models is enabled by a special feature of \textsc{JuMP}: the built-in \textsc{operator} function.
This feature detects and registers user-defined functions, facilitating their integration into optimization models.
However, to avoid the additional computational overhead of automatic differentiation (AD) \cite{Lubin2022_JuMP} and \cite{Revels2016}, \textit{explicit} gradients and Hessians (where applicable) are provided instead.

The reliability of the \softplus{} model is further improved through the \textsc{StatsFuns.jl} package\footnote{ \url{https://github.com/JuliaStats/StatsFuns.jl}}, and specifically the \textsc{log1pexp} and \textsc{logistic} functions designed to improve numerical stability by avoiding overflow/underflow.
A direct implementation of \eqref{softplus_VVC}-\eqref{softplus_VWC} is likely to cause numerical issues due to floating point overflow/underflow in intermediate evaluation steps, either in the function evaluation or in the derivative evaluation.
This is unsurprising since, for instance, as $\epsilon$ tends to 0, the exponential functions in \eqref{softplus_VVC} tend to infinity.

In each period, the quality of the solution $p^{*,\rm NLP}$ obtained from \nonsmooth{}, \softplus{}, or \spline{}, is compared to the solution $p^{*,\rm MINLP}$ obtained from \minlp{} by computing the \textit{relative} optimality gap defined as
\begin{equation}
	{\rm RelGap} = \left| \frac{p^{*,\rm MINLP}-{p^{*,\rm NLP}}}{p^{*,\rm MINLP}} \right|.
\end{equation}
When a solution from \minlp{} is not available, due to its intractability, the relative optimality gap is instead evaluated with respect to \nonsmooth{} as
\begin{equation}
	{\rm RelGap}= \left| \frac{p^{*, \rm nonsmooth}-p^{*, \rm smooth}}{p^{*, \rm nonsmooth}} \right|,
\end{equation}
where $p^{*, \rm smooth}$ is the objective function value from \softplus{} or \spline{} and $p^{*, \rm nonsmooth}$ from \nonsmooth{}.

\section{Numerical evaluation}\label{sec_numericaleval}

The \nonsmooth{}, \softplus{}, and \spline{} models are first validated against OpenDSS~\cite{Montenegro2022}, a widely used open-source DN simulation software with built-in incremental VVWC modeling capabilities as illustrated in Fig.~\ref{fig:loops}a.
This section also compares the accuracy, reliability, and computational performance of \minlp{}, \nonsmooth{}, \softplus{}, and \spline{}, and advises on the best choice of $\epsilon$ in the \softplus{} and \spline{} models that strikes a good tradeoff between accuracy, reliability, and computational efficiency.

\subsection{Validation against OpenDSS}\label{ValidationofVVWC}

Before proceeding with the benchmarking, it is crucial to highlight key differences between \textsc{OpenDSS} and the proposed optimization-based approaches, particularly in the modeling of the VVWC functions and the assumptions on voltage and current bounds.

First, as by default OpenDSS uses phase-to-ground voltages as inputs for its VVWC functions, phase-to-ground voltages are also used as inputs for the VVWC for the optimization-based methods.\footnote{To incorporate phase-to-neutral voltages in \textsc{OpenDSS}, lossless transformers are added between the corresponding buses and inverters.}

Second, being a simulation-only implementation, \textsc{OpenDSS} does not impose voltage bounds or any other inequality constraints.
Constraints \eqref{pg_bounds}-\eqref{thermal_limit} were therefore removed during this validation to ensure a like-for-like comparison.
These bounds are then reintroduced in Sections~\ref{subsec_comparisonwithminlp} and \ref{subsec_impactofepsilon} to demonstrate the capabilities of the optimization-based models in enabling many additional applications such as the ones described in Section~\ref{subsec_introduction}, which are not possible with simulation-only DN tools.

To enable a fair comparison with \textsc{Ipopt}, which is used with a default optimality tolerance of 1E-8, \textsc{OpenDSS} is invoked with control loop tolerances \texttt{VarChangeTolerance=1E-6} and \texttt{VoltageChangeTolerance=1E-7}.
The default value of \texttt{Tolerance=1E-4} is used for the PF tolerance.
\textsc{OpenDSS}'s VVC and VWC step size adjustments were set to \texttt{deltaQ\_factor=0.2} and \texttt{deltaP\_factor=0.5} to improve numerical stability.
Default values for these parameters, which are often set automatically by \textsc{OpenDSS} (typically \texttt{deltaQ\_factor=0.7} and \texttt{deltaP\_factor=1}), resulted in convergence failure in many cases, particularly on  \largenw{}.
The solution quality and computational performance of \textsc{OpenDSS} are compared with those of the proposed \nonsmooth{}, \softplus{}, and \spline{} formulations for the \largenw{} in Tables \ref{tab:539bus_comparison_with_opendss_optimality} and \ref{tab:539bus_comparison_with_opendss_solvetime}.
Results for \smallnw{} and \mediumnw{} are shown in Tables~\ref{tab:30bus_high_comparison_with_opendss_optimality}-\ref{tab:30bus_high_comparison_with_opendss_solvetime} and \ref{tab:99bus_high_comparison_with_opendss_optimality}-\ref{tab:99bus_high_comparison_with_opendss_solvetime}, respectively.

\begin{table*}[t!]
	\centering
	\caption{Comparison of the optimality of \textsc{OpenDSS}, \nonsmooth{}, \softplus{}, and \spline{} approximations across all time steps on the \smallnw{} system under the High PV adoption scenario, with phase-to-ground voltages as inputs to the VVWC (MA27, $\epsilon =10^{-8}$, max. ${\rm RelGap}$ = 2.35E-14 \%).}
	\begin{tabular}{r rrrr rrrr c c}
		\toprule
		\textbf{Time} &
		\multicolumn{4}{c}{\textbf{Objective value (kW)}} &
		\multicolumn{4}{c}{\textbf{Max VVWC error (\%)}} &
		\textbf{Dispatched} & \textbf{VUF} \\
		\cmidrule(l){2-5} \cmidrule(l){6-9}
		& \textsc{OpenDSS} & \nonsmooth & \softplus{} & \spline{}
		& \textsc{OpenDSS} & \nonsmooth & \softplus{} & \spline{}
		& \textbf{PV (kW)} & \textbf{(\%)} \\
		\midrule
		08:00 & 15.24   & 15.24   & 15.24   & 15.24   & 8.91E-05 & 3.75E-19 & 1.27E-14 & 3.75E-19 & 0.00   & 0.36 \\
		08:30 & -5.77   & -5.77   & -5.77   & -5.77   & 8.91E-05 & 5.61E-13 & 5.82E-13 & 5.61E-13 & 19.36  & 0.24 \\
		09:00 & -26.61  & -26.61  & -26.61  & -26.61  & 8.91E-05 & 5.58E-13 & 5.73E-13 & 5.58E-13 & 38.55  & 0.32 \\
		09:30 & -46.28  & -46.28  & -46.28  & -46.28  & 8.91E-05 & 5.58E-13 & 5.69E-13 & 5.58E-13 & 56.80  & 0.54 \\
		10:00 & -64.47  & -64.47  & -64.47  & -64.47  & 8.91E-05 & 5.63E-13 & 5.63E-13 & 5.63E-13 & 74.52  & 0.77 \\
		10:30 & -80.27  & -80.27  & -80.27  & -80.27  & 1.15E-04 & 1.53E-12 & 1.53E-12 & 1.53E-12 & 90.68  & 0.88 \\
		11:00 & -93.61  & -93.61  & -93.61  & -93.61  & 1.39E-04 & 1.45E-12 & 1.45E-12 & 1.45E-12 & 105.00 & 0.95 \\
		11:30 & -104.55 & -104.55 & -104.55 & -104.55 & 1.27E-04 & 1.41E-12 & 1.41E-12 & 1.41E-12 & 117.18 & 1.07 \\
		12:00 & -113.85 & -113.85 & -113.85 & -113.85 & 1.20E-04 & 1.50E-12 & 1.50E-12 & 1.50E-12 & 127.90 & 1.15 \\
		12:30 & -120.88 & -120.88 & -120.88 & -120.88 & 1.31E-04 & 4.42E-13 & 4.42E-13 & 4.42E-13 & 136.37 & 1.20 \\
		13:00 & -124.11 & -124.11 & -124.11 & -124.11 & 1.43E-04 & 1.49E-12 & 1.44E-12 & 1.49E-12 & 140.26 & 1.24 \\
		13:30 & -125.66 & -125.66 & -125.66 & -125.66 & 1.32E-04 & 1.40E-12 & 1.42E-12 & 1.40E-12 & 142.34 & 1.26 \\
		14:00 & -125.35 & -125.35 & -125.35 & -125.35 & 1.17E-04 & 1.41E-12 & 1.42E-12 & 1.41E-12 & 142.39 & 1.29 \\
		14:30 & -124.55 & -124.55 & -124.55 & -124.55 & 1.05E-04 & 1.43E-12 & 1.43E-12 & 1.43E-12 & 142.04 & 1.22 \\
		15:00 & -122.16 & -122.16 & -122.16 & -122.16 & 1.06E-04 & 4.34E-13 & 4.60E-13 & 4.34E-13 & 140.22 & 1.16 \\
		15:30 & -117.41 & -117.41 & -117.41 & -117.41 & 1.50E-04 & 1.40E-12 & 1.38E-12 & 1.40E-12 & 136.12 & 1.22 \\
		16:00 & -107.90 & -107.90 & -107.90 & -107.90 & 1.22E-04 & 1.38E-12 & 1.38E-12 & 1.38E-12 & 127.25 & 1.27 \\
		16:30 & -96.52  & -96.52  & -96.52  & -96.52  & 1.63E-04 & 1.41E-12 & 1.40E-12 & 1.41E-12 & 116.82 & 1.05 \\
		17:00 & -83.17  & -83.17  & -83.17  & -83.17  & 1.47E-04 & 1.26E-12 & 1.25E-12 & 1.26E-12 & 104.64 & 0.89 \\
		17:30 & -66.67  & -66.67  & -66.67  & -66.67  & 8.91E-05 & 5.54E-13 & 5.54E-13 & 5.54E-13 & 89.72  & 0.82 \\
		18:00 & -49.51  & -49.51  & -49.51  & -49.51  & 8.91E-05 & 5.67E-13 & 5.71E-13 & 5.67E-13 & 73.30  & 0.59 \\
		18:30 & -31.52  & -31.52  & -31.52  & -31.52  & 8.91E-05 & 5.73E-13 & 5.71E-13 & 5.73E-13 & 56.10  & 0.49 \\
		19:00 & -12.79  & -12.79  & -12.79  & -12.79  & 8.91E-05 & 5.73E-13 & 5.84E-13 & 5.73E-13 & 37.78  & 0.46 \\
		19:30 & 5.30    & 5.30    & 5.30    & 5.30    & 8.91E-05 & 5.68E-13 & 5.75E-13 & 5.68E-13 & 19.05  & 0.44 \\
		20:00 & 22.08   & 22.08   & 22.08   & 22.08   & 8.91E-05 & 3.55E-19 & 8.46E-15 & 3.55E-19 & 0.00   & 0.51                 \\
		\bottomrule
	\end{tabular}
	\label{tab:30bus_high_comparison_with_opendss_optimality}
\end{table*}

\begin{table*}[h!]
	\centering
	\caption{Computation time and iteration counts for \textsc{OpenDSS}, \nonsmooth{}, \softplus{}, and \spline{} approximations across all time steps on the \smallnw{} system under the High PV adoption scenario, with phase-to-ground voltages as inputs to the VVWC (MA27, $\epsilon =10^{-8}$).}
	\begin{tabular}{r rrrr rrrr}
		\toprule
		\textbf{Time} &
		\multicolumn{4}{c}{\textbf{Computation time (s)}} &
		\multicolumn{4}{c}{\textbf{Iterations (-)}} \\
		\cmidrule(l){2-5} \cmidrule(l){6-9}
		& \textsc{OpenDSS} & \nonsmooth & \softplus{} & \spline{} 
		& \textsc{OpenDSS} & \nonsmooth & \softplus{} & \spline{} \\
		\midrule
		08:00 & 1.85E-03 & 3.10E-02 & 3.00E-02 & 3.90E-02 & 118 & 19 & 19 & 19 \\
		08:30 & 1.89E-03 & 1.90E-02 & 1.70E-02 & 1.90E-02 & 129 & 9  & 9  & 9  \\
		09:00 & 2.04E-03 & 1.60E-02 & 1.70E-02 & 1.90E-02 & 143 & 9  & 9  & 9  \\
		09:30 & 2.01E-03 & 1.60E-02 & 1.70E-02 & 1.90E-02 & 146 & 9  & 9  & 9  \\
		10:00 & 2.04E-03 & 1.60E-02 & 1.70E-02 & 2.10E-02 & 154 & 9  & 9  & 9  \\
		10:30 & 2.06E-03 & 1.80E-02 & 1.70E-02 & 2.10E-02 & 160 & 9  & 9  & 9  \\
		11:00 & 2.00E-03 & 1.70E-02 & 1.80E-02 & 2.00E-02 & 182 & 10 & 10 & 10 \\
		11:30 & 2.05E-03 & 1.60E-02 & 1.70E-02 & 2.00E-02 & 176 & 9  & 9  & 9  \\
		12:00 & 2.26E-03 & 1.60E-02 & 1.80E-02 & 2.10E-02 & 211 & 10 & 10 & 10 \\
		12:30 & 1.89E-03 & 2.50E-02 & 2.20E-02 & 3.20E-02 & 173 & 14 & 14 & 14 \\
		13:00 & 1.87E-03 & 2.50E-02 & 2.20E-02 & 2.40E-02 & 172 & 13 & 13 & 13 \\
		13:30 & 1.86E-03 & 2.40E-02 & 2.10E-02 & 2.60E-02 & 169 & 13 & 13 & 13 \\
		14:00 & 1.88E-03 & 2.10E-02 & 2.20E-02 & 2.40E-02 & 168 & 13 & 13 & 13 \\
		14:30 & 1.88E-03 & 2.10E-02 & 2.30E-02 & 2.40E-02 & 163 & 13 & 13 & 13 \\
		15:00 & 1.88E-03 & 2.00E-02 & 2.40E-02 & 2.70E-02 & 166 & 15 & 15 & 15 \\
		15:30 & 1.88E-03 & 2.10E-02 & 2.20E-02 & 2.60E-02 & 183 & 13 & 13 & 13 \\
		16:00 & 1.88E-03 & 1.70E-02 & 1.70E-02 & 2.10E-02 & 176 & 10 & 10 & 10 \\
		16:30 & 2.01E-03 & 1.90E-02 & 1.80E-02 & 2.00E-02 & 171 & 10 & 10 & 10 \\
		17:00 & 1.94E-03 & 1.60E-02 & 1.70E-02 & 2.00E-02 & 149 & 9  & 9  & 9  \\
		17:30 & 2.00E-03 & 1.70E-02 & 1.60E-02 & 1.90E-02 & 157 & 9  & 9  & 9  \\
		18:00 & 1.95E-03 & 1.80E-02 & 2.20E-02 & 1.90E-02 & 147 & 9  & 9  & 9  \\
		18:30 & 1.90E-03 & 1.70E-02 & 2.20E-02 & 2.00E-02 & 140 & 9  & 9  & 9  \\
		19:00 & 1.88E-03 & 1.70E-02 & 1.60E-02 & 1.90E-02 & 132 & 9  & 9  & 9  \\
		19:30 & 1.85E-03 & 1.60E-02 & 1.70E-02 & 1.80E-02 & 128 & 9  & 9  & 9  \\
		20:00 & 1.73E-03 & 2.90E-02 & 3.00E-02 & 3.50E-02 & 118 & 19 & 19 & 19 \\
		\bottomrule
	\end{tabular}
	\label{tab:30bus_high_comparison_with_opendss_solvetime}
\end{table*}

\begin{table*}[h!]
	\centering
	\caption{Comparison of the optimality of \textsc{OpenDSS}, \nonsmooth{}, \softplus{}, and \spline{} approximations across all time steps on the \mediumnw{} system under the High PV adoption scenario, with phase-to-ground voltages as inputs to the VVWC (MA27, $\epsilon =10^{-8}$, max. ${\rm RelGap}$ = 3.67E-14\%).}
	\begin{tabular}{r rrrr rrrr c}
		\toprule
		\textbf{Time} &
		\multicolumn{4}{c}{\textbf{Objective value (kW)}} &
		\multicolumn{4}{c}{\textbf{Max VVWC error (\%)}} &
		\textbf{Dispatched PV (kW)} \\
		\cmidrule(l){2-5} \cmidrule(l){6-9}
		& \textsc{OpenDSS} & \nonsmooth & \softplus{} & \spline{}
		& \textsc{OpenDSS} & \nonsmooth & \softplus{} & \spline{}
		& \nonsmooth \\
		\midrule
		Time & OpenDSS & Nonsmooth & Softplus & Spline & OpenDSS & Nonsmooth & Softplus & Spline &  \\
		08:00 & 31.39 & 31.39 & 31.39 & 31.39 & 8.91E-05 & 3.32E-17 & 1.21E-14 & 3.32E-17 & 0.00 \\
		08:30 & -4.84 & -4.84 & -4.84 & -4.84 & 8.91E-05 & 3.17E-15 & 2.09E-14 & 3.17E-15 & 32.94 \\
		09:00 & -41.59 & -41.59 & -41.59 & -41.59 & 8.91E-05 & 2.33E-14 & 3.35E-14 & 2.33E-14 & 65.32 \\
		09:30 & -74.77 & -74.77 & -74.77 & -74.77 & 8.92E-05 & 7.49E-12 & 7.48E-12 & 7.49E-12 & 96.28 \\
		10:00 & -104.95 & -104.95 & -104.95 & -104.95 & 1.34E-04 & 1.86E-13 & 1.73E-13 & 1.86E-13 & 125.91 \\
		10:30 & -130.91 & -130.91 & -130.91 & -130.91 & 1.50E-04 & 1.80E-13 & 1.84E-13 & 1.80E-13 & 153.12 \\
		11:00 & -152.77 & -152.77 & -152.77 & -152.77 & 1.65E-04 & 1.44E-13 & 1.44E-13 & 1.44E-13 & 177.55 \\
		11:30 & -170.65 & -170.65 & -170.65 & -170.65 & 1.46E-04 & 1.55E-11 & 1.54E-11 & 1.55E-11 & 198.96 \\
		12:00 & -184.69 & -184.69 & -184.69 & -184.69 & 1.40E-04 & 5.96E-12 & 5.96E-12 & 5.96E-12 & 216.97 \\
		12:30 & -194.97 & -194.97 & -194.97 & -194.97 & 1.56E-04 & 3.76E-13 & 3.93E-13 & 3.76E-13 & 229.51 \\
		13:00 & -199.57 & -199.57 & -199.57 & -199.57 & 1.57E-04 & 2.71E-13 & 2.88E-13 & 2.71E-13 & 236.02 \\
		13:30 & -201.18 & -201.18 & -201.18 & -201.18 & 1.36E-04 & 1.52E-13 & 1.56E-13 & 1.52E-13 & 238.69 \\
		14:00 & -200.34 & -200.34 & -200.34 & -200.34 & 1.27E-04 & 1.44E-13 & 1.18E-13 & 1.44E-13 & 238.68 \\
		14:30 & -199.12 & -199.12 & -199.12 & -199.12 & 1.41E-04 & 1.90E-13 & 1.74E-13 & 1.90E-13 & 238.10 \\
		15:00 & -196.04 & -196.04 & -196.04 & -196.04 & 1.35E-04 & 1.95E-13 & 2.16E-13 & 1.95E-13 & 236.04 \\
		15:30 & -188.66 & -188.66 & -188.66 & -188.66 & 1.36E-04 & 1.76E-08 & 1.76E-08 & 1.76E-08 & 229.82 \\
		16:00 & -174.37 & -174.37 & -174.37 & -174.37 & 1.49E-04 & 3.00E-13 & 3.26E-13 & 3.00E-13 & 216.24 \\
		16:30 & -154.40 & -154.40 & -154.40 & -154.40 & 1.40E-04 & 4.78E-13 & 4.78E-13 & 4.78E-13 & 198.00 \\
		17:00 & -132.55 & -132.55 & -132.55 & -132.55 & 1.41E-04 & 1.64E-11 & 1.64E-11 & 1.64E-11 & 177.00 \\
		17:30 & -105.72 & -105.72 & -105.72 & -105.72 & 1.32E-04 & 1.13E-11 & 1.13E-11 & 1.13E-11 & 152.51 \\
		18:00 & -77.34 & -77.34 & -77.34 & -77.34 & 8.92E-05 & 8.46E-15 & 2.11E-14 & 8.46E-15 & 125.24 \\
		18:30 & -45.21 & -45.21 & -45.21 & -45.21 & 8.91E-05 & 8.25E-14 & 7.40E-14 & 8.25E-14 & 96.11 \\
		19:00 & -13.60 & -13.60 & -13.60 & -13.60 & 8.91E-05 & 4.23E-15 & 1.90E-14 & 4.23E-15 & 65.05 \\
		19:30 & 17.15 & 17.15 & 17.15 & 17.15 & 8.91E-05 & 3.17E-15 & 2.01E-14 & 3.17E-15 & 32.79 \\
		20:00 & 46.73 & 46.73 & 46.73 & 46.73 & 9.01E-05 & 1.55E-15 & 6.21E-15 & 1.55E-15 & 0.00 \\
		\bottomrule
	\end{tabular}
	\label{tab:99bus_high_comparison_with_opendss_optimality}
\end{table*}

\begin{table*}[h!]
	\centering
	\caption{Computation time and iteration counts for \textsc{OpenDSS}, \nonsmooth{}, \softplus{}, and \spline{} approximations across all time steps on the \mediumnw{} system under the High PV adoption scenario, with phase-to-ground voltages as inputs to the VVWC (MA27, $\epsilon =10^{-8}$).}
	\begin{tabular}{r rrrr rrrr}
		\toprule
		\textbf{Time} &
		\multicolumn{4}{c}{\textbf{Computation time (s)}} &
		\multicolumn{4}{c}{\textbf{Iterations (-)}} \\
		\cmidrule(l){2-5} \cmidrule(l){6-9}
		& \textsc{OpenDSS} & \nonsmooth & \softplus{} & \spline{} 
		& \textsc{OpenDSS} & \nonsmooth & \softplus{} & \spline{} \\
		\midrule
		08:00 & 1.60E-02 & 4.70E-02 & 5.00E-02 & 5.10E-02 & 118 & 8 & 8 & 8 \\
		08:30 & 1.57E-02 & 4.90E-02 & 5.00E-02 & 5.00E-02 & 114 & 9 & 9 & 9 \\
		09:00 & 1.51E-02 & 4.70E-02 & 4.60E-02 & 5.00E-02 & 113 & 9 & 9 & 9 \\
		09:30 & 1.58E-02 & 4.70E-02 & 4.70E-02 & 4.60E-02 & 120 & 9 & 9 & 9 \\
		10:00 & 1.51E-02 & 5.50E-02 & 5.90E-02 & 5.80E-02 & 120 & 11 & 11 & 11 \\
		10:30 & 1.52E-02 & 5.60E-02 & 6.20E-02 & 5.80E-02 & 120 & 11 & 11 & 11 \\
		11:00 & 1.56E-02 & 5.70E-02 & 5.80E-02 & 5.80E-02 & 128 & 11 & 11 & 11 \\
		11:30 & 1.62E-02 & 5.40E-02 & 5.40E-02 & 5.40E-02 & 137 & 11 & 11 & 11 \\
		12:00 & 1.68E-02 & 6.00E-02 & 6.40E-02 & 6.20E-02 & 146 & 12 & 12 & 12 \\
		12:30 & 1.78E-02 & 6.40E-02 & 7.10E-02 & 6.80E-02 & 155 & 13 & 13 & 13 \\
		13:00 & 1.79E-02 & 7.30E-02 & 8.00E-02 & 7.60E-02 & 158 & 15 & 15 & 15 \\
		13:30 & 1.81E-02 & 7.60E-02 & 8.40E-02 & 8.20E-02 & 160 & 16 & 16 & 16 \\
		14:00 & 1.86E-02 & 7.30E-02 & 7.90E-02 & 8.60E-02 & 168 & 15 & 15 & 15 \\
		14:30 & 1.84E-02 & 8.00E-02 & 8.80E-02 & 8.30E-02 & 162 & 16 & 16 & 16 \\
		15:00 & 1.76E-02 & 7.70E-02 & 8.10E-02 & 7.90E-02 & 155 & 16 & 16 & 16 \\
		15:30 & 1.70E-02 & 7.90E-02 & 8.80E-02 & 8.40E-02 & 146 & 16 & 16 & 16 \\
		16:00 & 1.63E-02 & 6.00E-02 & 6.60E-02 & 6.60E-02 & 138 & 12 & 12 & 12 \\
		16:30 & 1.55E-02 & 5.60E-02 & 6.30E-02 & 6.30E-02 & 127 & 11 & 11 & 11 \\
		17:00 & 1.50E-02 & 6.60E-02 & 7.20E-02 & 7.10E-02 & 121 & 13 & 13 & 13 \\
		17:30 & 1.55E-02 & 5.40E-02 & 6.30E-02 & 5.80E-02 & 121 & 10 & 10 & 10 \\
		18:00 & 1.52E-02 & 5.40E-02 & 6.20E-02 & 5.80E-02 & 119 & 10 & 10 & 10 \\
		18:30 & 1.49E-02 & 4.70E-02 & 5.00E-02 & 5.40E-02 & 113 & 9 & 9 & 9 \\
		19:00 & 1.47E-02 & 4.90E-02 & 5.00E-02 & 5.40E-02 & 115 & 9 & 9 & 9 \\
		19:30 & 1.51E-02 & 4.90E-02 & 6.20E-02 & 7.10E-02 & 118 & 9 & 9 & 9 \\
		20:00 & 1.59E-02 & 4.50E-02 & 5.00E-02 & 5.40E-02 & 129 & 8 & 8 & 8 \\
		\bottomrule
	\end{tabular}
	\label{tab:99bus_high_comparison_with_opendss_solvetime}
\end{table*}

Two key observations can be drawn from these findings.
First, at times of the day before 10 AM and after 6 PM, \nonsmooth{}, \softplus{}, and \spline{} yield the exact \textit{same} solutions as \textsc{OpenDSS}, evidenced by similar objective function values and small VVWC errors.
Second, Table \ref{tab:539bus_comparison_with_opendss_solvetime} shows that, even under \texttt{MaxIterations=1000}, \textsc{OpenDSS} fails to converge in many instances with high PV generation between 10 AM and 6 PM on the \largenw{} under the High PV adoption Scenario.
In contrast, \nonsmooth{}, \softplus{}, and \spline{} consistently converge across all 25 snapshots for the three networks under every PV adoption scenario.
While \textsc{OpenDSS} experiences similar convergence issues under the \largenw{} Medium PV adoption scenario, it consistently converges for the \smallnw{} and \mediumnw{} under all scenarios, and for the \largenw{} under the Low PV adoption scenario.
This numerically corroborates the superior reliability of the proposed optimization-based models under the proposed implementation setup.
In fact, a more equitable comparison is attained by setting \textsc{OpenDSS}'s all tolerance parameters to 1E-8.
However, in doing so, \textsc{OpenDSS}'s convergence becomes significantly less robust, particularly when the power flow tolerance (\texttt{Tolerance}) is set below 1E-4.

More interestingly, the findings in Tables~\ref{tab:30bus_high_comparison_with_opendss_optimality}, \ref{tab:99bus_high_comparison_with_opendss_optimality}, and \ref{tab:539bus_comparison_with_opendss_optimality} show that, when it converges, \textsc{OpenDSS} yields the exact same solutions as the optimization-based counterparts.
This can be explained by VVWC being fundamentally a distributed control solution that minimizes active power curtailment, noting that the sensitivities $\Delta V/\Delta P$ are naturally higher for customers further from the feeder head, who therefore witness more curtailment.
A more general way to capture this is to cast the VVWO problem as a \textit{bilevel} problem, where the upper level represents a distribution utility’s or retailer’s objective to minimize power imports from the grid, while the lower level reflects end-users' objective to maximize PV exports.
However, under the specific objective function in \eqref{objective}, which maximizes PV generation, the bilevel problem is equivalent to the single-level problem in \eqref{VVWO}.
This explains the observed outcomes and clarifies the conditions under which non-incremental and incremental control strategies can yield similar results in the context of VVWC.

\begin{table*}[t!]
	\centering
	\caption{Comparison of the optimality of \textsc{OpenDSS}, \nonsmooth{}, \softplus{}, and \spline{} approximations across all time steps on the \largenw{} system under the High PV adoption scenario, with \eqref{pg_bounds}–\eqref{voltage_bounds} ignored and phase-to-ground voltages as inputs to the VVWC (MA27, $\epsilon =10^{-8}$, max. ${\rm RelGap}$ = 3.07E-08\%). DNC stands for "did not converge".}
	\begin{tabular}{r rrrr rrrr c c}
		\toprule
		\textbf{Time} &
		\multicolumn{4}{c}{\textbf{Objective value (kW)}} &
		\multicolumn{4}{c}{\textbf{Max VVWC error (\%)}} &
		\textbf{Dispatched} & \textbf{VUF} \\
		\cmidrule(l){2-5} \cmidrule(l){6-9}
		& \textsc{OpenDSS} & \nonsmooth & \softplus{} & \spline{}
		& \textsc{OpenDSS} & \nonsmooth & \softplus{} & \spline{}
		& \textbf{PV (kW)} & \textbf{(\%)} \\
		\midrule
		08:00 & 166.20  & 166.20   & 166.20   & 166.20   & 1.55E-04 & 5.33E-13 & 4.98E-13 & 5.33E-13 & 0.00    & 0.59 \\
		08:30 & -60.53  & -60.53   & -60.53   & -60.53   & 8.92E-05 & 5.06E-12 & 5.06E-12 & 5.06E-12 & 206.36  & 0.57 \\
		09:00 & -274.22 & -274.22  & -274.22  & -274.22  & 1.71E-04 & 4.66E-11 & 4.68E-11 & 4.66E-11 & 409.23  & 1.26 \\
		09:30 & -459.78 & -459.78  & -459.78  & -459.78  & 1.70E-04 & 5.67E-11 & 5.67E-11 & 5.67E-11 & 604.70  & 2.49 \\
		10:00 & DNC     & -617.57  & -617.57  & -617.57  & DNC      & 6.26E-11 & 6.26E-11 & 6.26E-11 & 790.26  & 3.16 \\
		10:30 & DNC     & -747.96  & -747.96  & -747.96  & DNC      & 6.92E-11 & 6.92E-11 & 6.92E-11 & 962.09  & 3.82 \\
		11:00 & DNC     & -849.40  & -849.40  & -849.40  & DNC      & 1.01E-07 & 1.01E-07 & 1.01E-07 & 1106.76 & 4.45 \\
		11:30 & DNC     & -922.46  & -922.46  & -922.46  & DNC      & 8.94E-08 & 8.94E-08 & 8.94E-08 & 1218.50 & 5.03 \\
		12:00 & DNC     & -973.76  & -973.76  & -973.76  & DNC      & 6.42E-08 & 6.42E-08 & 6.42E-08 & 1299.03 & 5.56 \\
		12:30 & DNC     & -1005.85 & -1005.85 & -1005.85 & DNC      & 5.29E-08 & 5.29E-08 & 5.29E-08 & 1345.58 & 5.70 \\
		13:00 & DNC     & -1020.82 & -1020.82 & -1020.82 & DNC      & 5.27E-08 & 5.27E-08 & 5.27E-08 & 1368.72 & 5.77 \\
		13:30 & DNC     & -1026.86 & -1026.86 & -1026.86 & DNC      & 5.27E-08 & 5.27E-08 & 5.27E-08 & 1380.18 & 5.82 \\
		14:00 & DNC     & -1028.33 & -1028.33 & -1028.33 & DNC      & 5.22E-08 & 5.22E-08 & 5.22E-08 & 1386.00 & 5.84 \\
		14:30 & DNC     & -1021.52 & -1021.52 & -1021.52 & DNC      & 4.68E-08 & 4.68E-08 & 4.68E-08 & 1384.12 & 5.80 \\
		15:00 & DNC     & -1007.61 & -1007.61 & -1007.61 & DNC      & 4.59E-08 & 4.59E-08 & 4.59E-08 & 1376.02 & 5.74 \\
		15:30 & DNC     & -984.06  & -984.06  & -984.06  & DNC      & 5.40E-08 & 5.40E-08 & 5.40E-08 & 1359.20 & 5.60 \\
		16:00 & DNC     & -937.62  & -937.62  & -937.62  & DNC      & 1.05E-07 & 1.05E-07 & 1.05E-07 & 1312.62 & 5.29 \\
		16:30 & DNC     & -868.10  & -868.10  & -868.10  & DNC      & 1.01E-07 & 1.01E-07 & 1.01E-07 & 1231.71 & 4.66 \\
		17:00 & DNC     & -770.65  & -770.65  & -770.65  & DNC      & 7.69E-08 & 7.69E-08 & 7.69E-08 & 1115.92 & 4.12 \\
		17:30 & DNC     & -641.07  & -641.07  & -641.07  & DNC      & 6.59E-11 & 6.59E-11 & 6.59E-11 & 961.73  & 3.50 \\
		18:00 & DNC     & -492.16  & -492.16  & -492.16  & DNC      & 5.79E-11 & 5.79E-11 & 5.79E-11 & 789.79  & 2.80 \\
		18:30 & -327.06 & -327.06  & -327.06  & -327.06  & 1.67E-04 & 5.01E-11 & 5.01E-11 & 5.01E-11 & 604.79  & 1.76 \\
		19:00 & -140.53 & -140.53  & -140.53  & -140.53  & 8.97E-05 & 4.80E-12 & 4.81E-12 & 4.80E-12 & 408.88  & 1.01 \\
		19:30 & 46.82   & 46.82    & 46.82    & 46.82    & 9.88E-05 & 4.57E-12 & 4.57E-12 & 4.57E-12 & 206.23  & 0.78 \\
		20:00 & 241.90  & 241.90   & 241.90   & 241.90   & 1.54E-04 & 5.12E-13 & 5.09E-13 & 5.12E-13 & 0.00    & 0.94 \\
		\bottomrule
	\end{tabular}
	\label{tab:539bus_comparison_with_opendss_optimality}
\end{table*}

\begin{table*}[t!]
	\centering
	\caption{Computation time and iteration counts for \textsc{OpenDSS}, \nonsmooth{}, \softplus{}, and \spline{} approximations across all time steps on the \largenw{} system under the High PV adoption scenario, with \eqref{pg_bounds}–\eqref{voltage_bounds} ignored and phase-to-ground voltages as inputs to the VVWC (MA27, $\epsilon =10^{-8}$).}
	\begin{tabular}{r rrrr rrrr}
		\toprule
		\textbf{Time} &
		\multicolumn{4}{c}{\textbf{Computation time (s)}} &
		\multicolumn{4}{c}{\textbf{Iterations (-)}} \\
		\cmidrule(l){2-5} \cmidrule(l){6-9}
		& \textsc{OpenDSS} & \nonsmooth & \softplus{} & \spline{}
		& \textsc{OpenDSS} & \nonsmooth & \softplus{} & \spline{} \\
		\midrule
		08:00 & 0.18 & 0.64 & 0.63 & 0.69 & 132 & 19 & 19 & 19 \\
		08:30 & 0.18 & 0.38 & 0.39 & 0.41 & 124 & 10 & 10 & 10 \\
		09:00 & 0.19 & 0.39 & 0.39 & 0.42 & 159 & 10 & 10 & 10 \\
		09:30 & 0.20 & 0.42 & 0.46 & 0.47 & 181 & 11 & 11 & 11 \\
		10:00 & DNC  & 0.46 & 0.44 & 0.46 & DNC & 11 & 11 & 11 \\
		10:30 & DNC  & 0.51 & 0.53 & 0.52 & DNC & 14 & 14 & 14 \\
		11:00 & DNC  & 0.62 & 0.71 & 0.65 & DNC & 18 & 18 & 18 \\
		11:30 & DNC  & 0.60 & 0.58 & 0.75 & DNC & 17 & 17 & 17 \\
		12:00 & DNC  & 0.84 & 0.70 & 0.77 & DNC & 21 & 21 & 21 \\
		12:30 & DNC  & 0.76 & 0.73 & 0.87 & DNC & 22 & 22 & 22 \\
		13:00 & DNC  & 0.74 & 0.77 & 0.85 & DNC & 22 & 22 & 22 \\
		13:30 & DNC  & 0.78 & 0.85 & 1.07 & DNC & 23 & 23 & 23 \\
		14:00 & DNC  & 0.73 & 0.90 & 0.79 & DNC & 22 & 22 & 22 \\
		14:30 & DNC  & 0.73 & 0.93 & 0.86 & DNC & 22 & 22 & 22 \\
		15:00 & DNC  & 0.71 & 0.75 & 0.82 & DNC & 21 & 21 & 21 \\
		15:30 & DNC  & 0.74 & 0.75 & 0.79 & DNC & 22 & 22 & 22 \\
		16:00 & DNC  & 0.62 & 0.62 & 0.69 & DNC & 18 & 18 & 18 \\
		16:30 & DNC  & 0.58 & 0.61 & 0.63 & DNC & 17 & 17 & 17 \\
		17:00 & DNC  & 0.57 & 0.59 & 0.62 & DNC & 16 & 16 & 16 \\
		17:30 & DNC  & 0.48 & 0.51 & 0.51 & DNC & 13 & 13 & 13 \\
		18:00 & DNC  & 0.43 & 0.47 & 0.45 & DNC & 11 & 11 & 11 \\
		18:30 & 0.19 & 0.42 & 0.46 & 0.44 & 154 & 11 & 11 & 11 \\
		19:00 & 0.18 & 0.38 & 0.37 & 0.42 & 133 & 9  & 9  & 9  \\
		19:30 & 0.17 & 0.42 & 0.39 & 0.46 & 125 & 10 & 10 & 10 \\
		20:00 & 0.17 & 0.65 & 0.65 & 0.70 & 140 & 19 & 19 & 19 \\
		\bottomrule
	\end{tabular}
	\label{tab:539bus_comparison_with_opendss_solvetime}
\end{table*}

\subsection{Comparison with \minlp{} } \label{subsec_comparisonwithminlp}

In this subsection, with the aim of evaluating capabilities unique to optimization-based models, the voltage bounds \eqref{voltage_bounds} of 0.9 and 1.1217 pu (258 V / 230 V) are reintroduced, and phase-to-neutral voltages are used as inputs to the VVWC.
Additionally, a neutral-to-ground voltage limit of 0.1 pu is imposed, a feat unique to four-wire modelling.
The accuracy and computational performance of \nonsmooth{}, \softplus{}, and \spline{} relative to \minlp{} are listed in Table \ref{tab:30bus_comparison_with_MINLP} for \smallnw{} under the High PV adoption scenario.
The following findings can be drawn from Table \ref{tab:30bus_comparison_with_MINLP}.
First, all four models, \minlp{}, \nonsmooth{}, \softplus{}, and \spline{}, converge to the exact same (feasible) solutions, as evidenced by a very small RelGap (maximum RelGap on the order of \num{e-7}).
Second, the \minlp{} model is, on average, \textit{one order of magnitude slower} than its NLP counterparts.
Third, the \softplus{}, and \spline{} models, even under the most stringent choice of $\epsilon =10^{-8}$, exhibit comparable computational performance compared to \nonsmooth{}.

On \mediumnw{}, which is accompanied by a greater number of VVWC constraints, and therefore more binary variables, the \minlp{} consistently fails to converge within 2 hours.
On \largenw{}, the \minlp{} model consistently fails to converge within 24 hours.
Consequently, the analysis in the next subsection focuses solely on \nonsmooth{}, \softplus{}, and \spline{} across all three networks and PV adoption levels.

\begin{table*}[t!]
	\centering
	\caption{Objective values, relative gaps, and computation times for \minlp{}, \nonsmooth{}, \softplus{}, and \spline{} on the \smallnw{} system under the High PV adoption scenario, with all inequality constraints enforced and phase-to-neutral voltages as inputs to the VVWC (MA27, $\epsilon =10^{-8}$).}
	\begin{tabular}{r c rrr rrrr}
		\toprule
		\textbf{Time} &
		\multicolumn{1}{c}{\textbf{Objective (kW)}} &
		\multicolumn{3}{c}{\textbf{RelGap (\%)}} &
		\multicolumn{4}{c}{\textbf{Computation time (s)}} \\
		\cmidrule(l){2-2} \cmidrule(l){3-5} \cmidrule(l){6-9}
		& \minlp &
		\nonsmooth & \softplus{} & \spline{} &
		\minlp & \nonsmooth & \softplus{} & \spline{} \\
		\midrule
		08:00 & 15.24   & 4.06E-07 & 4.06E-07 & 4.06E-07 & 134.93 & 0.03 & 0.04 & 0.04 \\
		08:30 & -5.77   & 1.02E-06 & 1.02E-06 & 1.02E-06 & 119.55 & 0.02 & 0.02 & 0.02 \\
		09:00 & -26.61  & 2.30E-07 & 2.30E-07 & 2.30E-07 & 127.09 & 0.02 & 0.02 & 0.02 \\
		09:30 & -46.28  & 1.45E-07 & 1.45E-07 & 1.45E-07 & 114.04 & 0.02 & 0.02 & 0.02 \\
		10:00 & -64.44  & 1.19E-07 & 1.19E-07 & 1.19E-07 & 172.65 & 0.02 & 0.02 & 0.02 \\
		10:30 & -80.17  & 8.85E-08 & 8.85E-08 & 8.85E-08 & 253.63 & 0.02 & 0.02 & 0.02 \\
		11:00 & -93.41  & 1.40E-07 & 1.40E-07 & 1.40E-07 & 178.56 & 0.02 & 0.02 & 0.02 \\
		11:30 & -104.24 & 3.29E-08 & 3.29E-08 & 3.29E-08 & 142.71 & 0.02 & 0.02 & 0.02 \\
		12:00 & -113.42 & 1.42E-07 & 1.42E-07 & 1.42E-07 & 138.14 & 0.02 & 0.02 & 0.03 \\
		12:30 & -119.63 & 5.62E-07 & 5.62E-07 & 5.62E-07 & 140.35 & 0.03 & 0.03 & 0.04 \\
		13:00 & -122.39 & 5.03E-07 & 5.03E-07 & 5.03E-07 & 154.95 & 0.03 & 0.02 & 0.03 \\
		13:30 & -123.72 & 6.73E-07 & 6.73E-07 & 6.73E-07 & 151.72 & 0.03 & 0.03 & 0.03 \\
		14:00 & -123.25 & 9.83E-07 & 9.83E-07 & 9.83E-07 & 178.96 & 0.02 & 0.03 & 0.03 \\
		14:30 & -122.60 & 6.45E-07 & 6.45E-07 & 6.45E-07 & 141.18 & 0.03 & 0.02 & 0.03 \\
		15:00 & -120.75 & 6.93E-07 & 6.93E-07 & 6.93E-07 & 136.15 & 0.03 & 0.03 & 0.03 \\
		15:30 & -116.41 & 5.25E-07 & 5.25E-07 & 5.25E-07 & 143.34 & 0.03 & 0.03 & 0.03 \\
		16:00 & -107.51 & 1.64E-07 & 1.64E-07 & 1.64E-07 & 131.96 & 0.02 & 0.02 & 0.03 \\
		16:30 & -96.25  & 1.23E-07 & 1.23E-07 & 1.23E-07 & 147.79 & 0.03 & 0.02 & 0.02 \\
		17:00 & -83.02  & 1.46E-07 & 1.46E-07 & 1.46E-07 & 159.98 & 0.02 & 0.02 & 0.02 \\
		17:30 & -66.64  & 1.04E-07 & 1.04E-07 & 1.04E-07 & 245.02 & 0.02 & 0.02 & 0.02 \\
		18:00 & -49.51  & 1.28E-07 & 1.28E-07 & 1.28E-07 & 199.38 & 0.02 & 0.03 & 0.03 \\
		18:30 & -31.52  & 1.99E-07 & 1.99E-07 & 1.99E-07 & 153.65 & 0.02 & 0.02 & 0.02 \\
		19:00 & -12.79  & 3.84E-07 & 3.84E-07 & 3.84E-07 & 140.93 & 0.02 & 0.02 & 0.02 \\
		19:30 & 5.30    & 9.28E-07 & 9.28E-07 & 9.28E-07 & 140.13 & 0.02 & 0.02 & 0.02 \\
		20:00 & 22.08   & 4.83E-07 & 4.83E-07 & 4.83E-07 & 141.05 & 0.03 & 0.04 & 0.04 \\
		\bottomrule
	\end{tabular}
	\label{tab:30bus_comparison_with_MINLP}
\end{table*}

\subsection{Accuracy, computational performance, and impact of the smoothness parameter} \label{subsec_impactofepsilon}

\begin{figure}[!h]
	\centering
	\begin{subfigure}{1\columnwidth}
		\includegraphics[width=1\columnwidth]{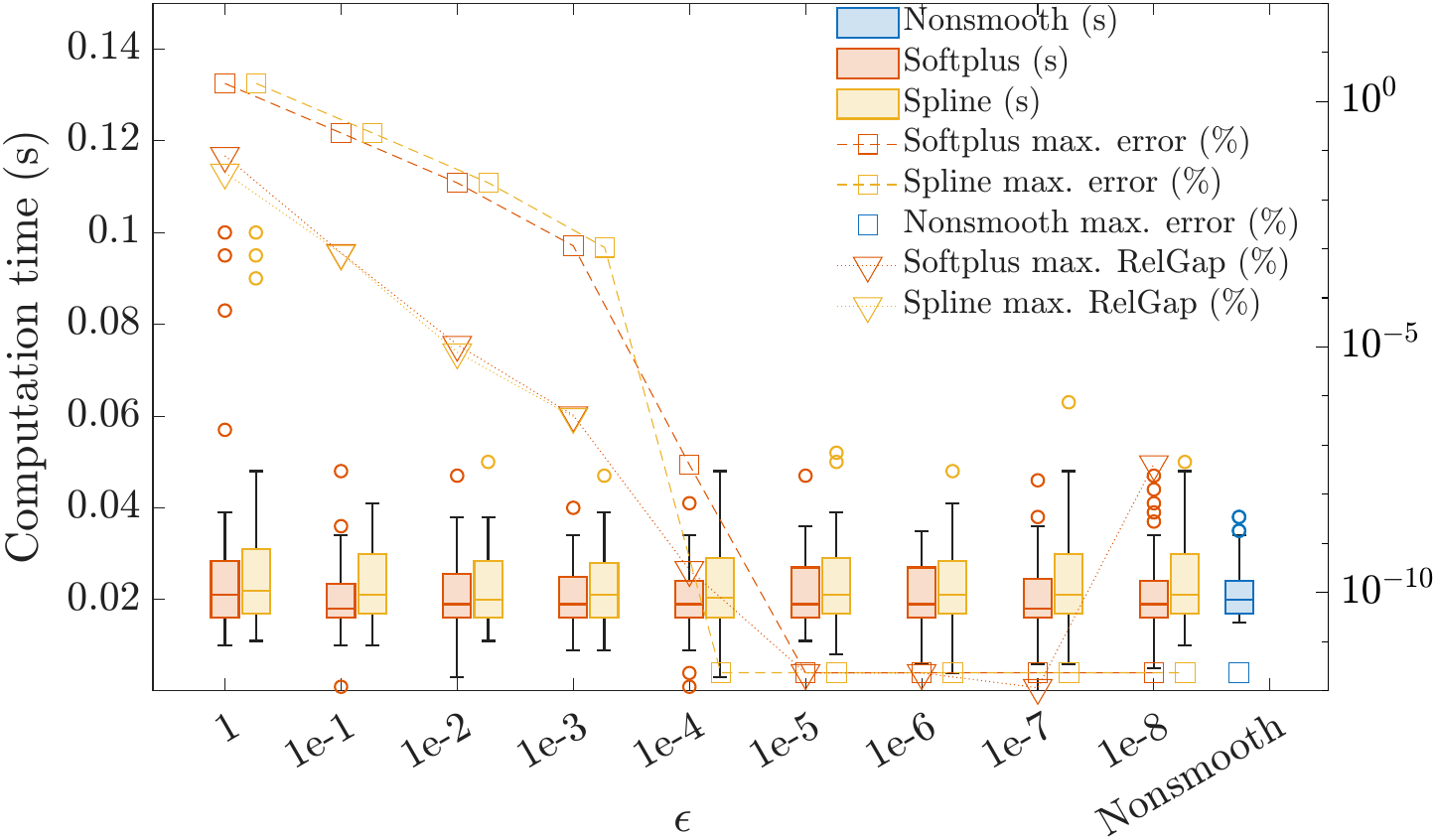}
		\caption{\smallnw{}}
		\label{fig_comprehensivecomparison_30bus}
	\end{subfigure}
	\hfill
	\begin{subfigure}{1\columnwidth}
		\includegraphics[width=1\columnwidth]{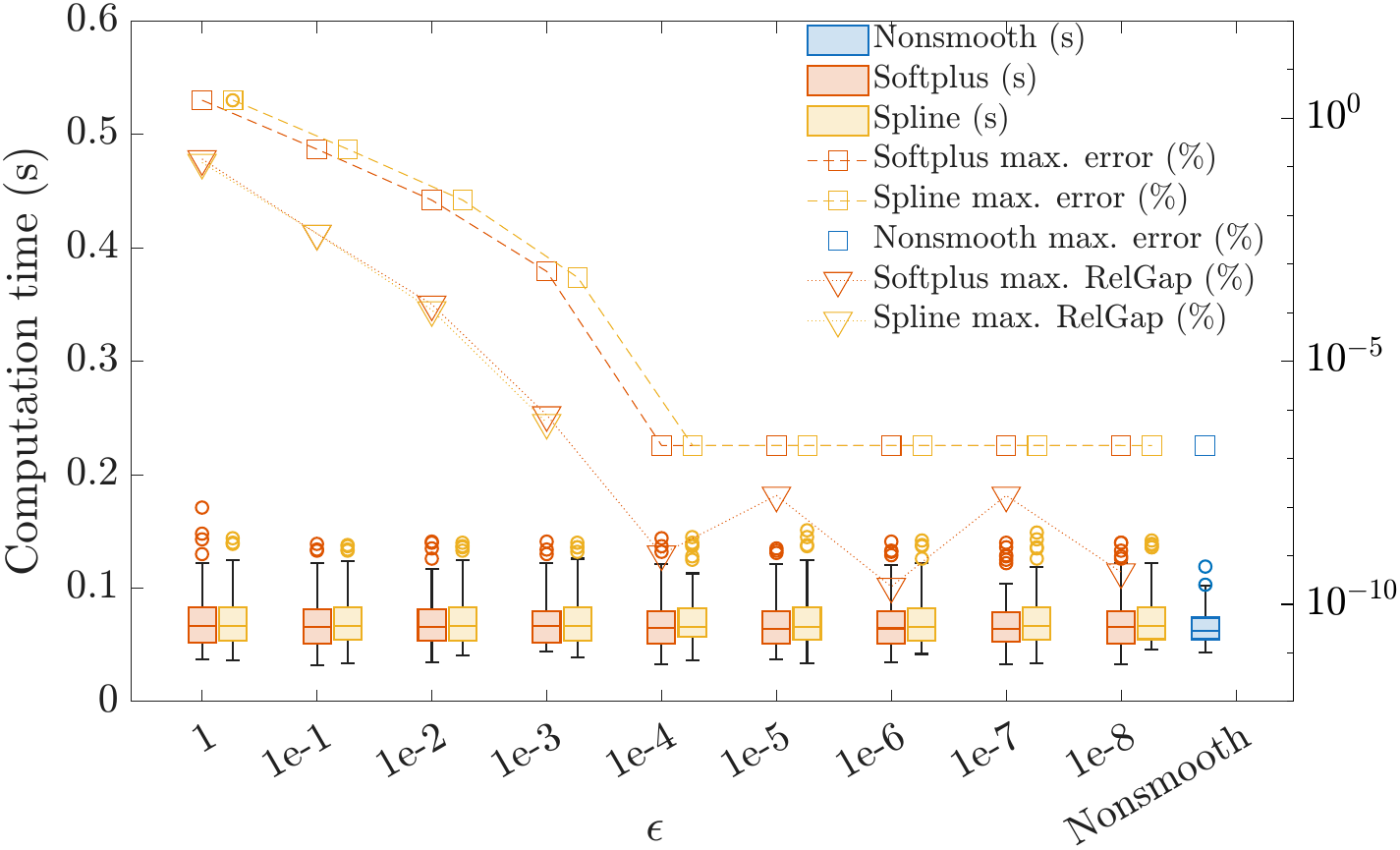}
		\caption{\mediumnw{}}
		\label{fig_comprehensivecomparison_99bus}
	\end{subfigure}
	\begin{subfigure}{1\columnwidth}
		\includegraphics[width=1\columnwidth]{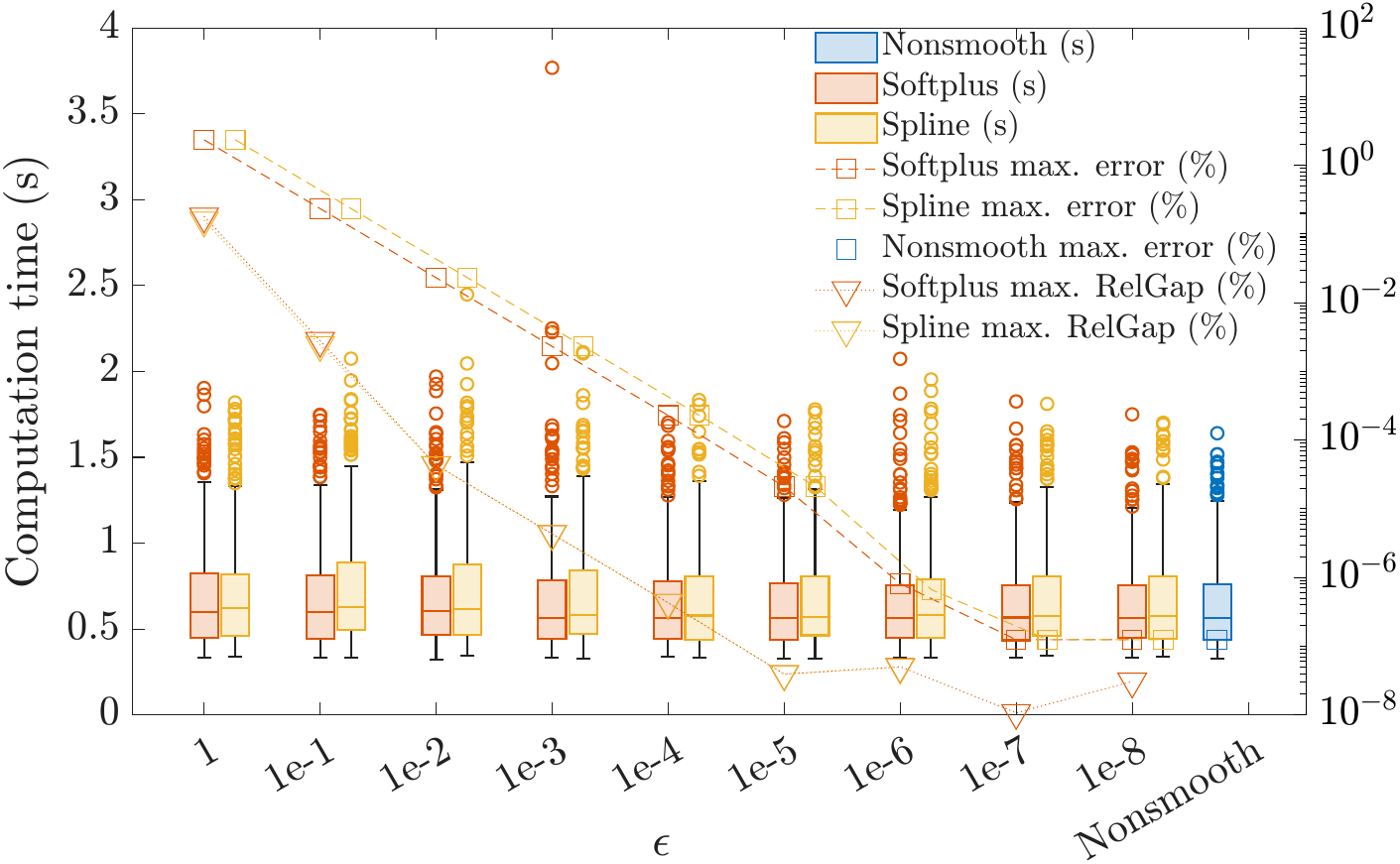}
		\caption{\largenw{}}
		\label{fig_comprehensivecomparison_539bus}
	\end{subfigure}
	\caption{Computation time, maximum VVWC error, and maximum RelGap of \nonsmooth{}, \softplus{}, and \spline{} on (a) \smallnw{}, (b) \mediumnw{}, and (c) \largenw{}. The approximation accuracy $\epsilon$ in \softplus{} and \spline{} is varied from $1$ to $10^{-8}$, for a total of 8,550 instances, including the 450 \nonsmooth{} instances.}
	\label{fig_comprehensivecomparison}
\end{figure}

Because \nonsmooth{} is now the only remaining exact model, the accuracy and computational performance of \softplus{} and \spline{} are now evaluated against \nonsmooth{} as the benchmark.
A rigorous evaluation of accuracy and computational performance of \nonsmooth{}, \softplus{}, and \spline{} on all three LV networks, under all three PV adoption scenarios, across all time steps, and for both phase-to-ground (without inequality constraints) and phase-to-neutral (with inequality constraints), is summarized in Fig. \ref{fig_comprehensivecomparison}.
Fig. \ref{fig_comprehensivecomparison} also shows a systematic assessment of the impact of the smoothness parameter $\epsilon$ in \softplus{} and \spline{} on accuracy and computational performance.
A total of 8,550 instances were evaluated, with all three methods converging across every instance.

Two keys observations can be made from Fig. \ref{fig_comprehensivecomparison}.
First, \softplus{} and \spline{} with $\epsilon = 10^{-8}$ converge to the same feasible, high-quality solutions across \textit{all} 900 instances, a testament to the high accuracy of the approximation.
This includes High PV adoption scenarios, which introduce a greater number of VVWC constraints.
The quality of the solution is measured by the maximum VVWC error (see \eqref{softplus_VVC}–\eqref{softplus_VWC} in \softplus{} and \eqref{eq_Qspline}–\eqref{eq_Pspline} in \spline{}) and the ${\rm RelGap}$.

Second, \nonsmooth{} exhibited the best overall computational performance across all three LV networks.
This is because gradient and Hessian function evaluations in \nonsmooth{} are computationally cheaper than their \softplus{} and \spline{} counterparts.
Interestingly, \nonsmooth{} converged for all 450 instances, suggesting that none of the solutions had voltages at or in close proximity to the breakpoints.
In rare cases where the solution has voltages at or in close proximity to the breakpoints, the line search step size becomes more sensitive to abrupt changes in the gradient, slowing convergence or hindering it altogether.
As an example of such case, consider the following simple problem
\begin{subequations} \label{VVCexample}
	\begin{align}
		{\mbox{minimize}} \quad & Q + U^{\rm m} \\
		\text{subject to} \quad & \text{\eqref{VVC_constraint}} , \\
		1.1 &\leq U^{\rm m} \leq 1.15 , \\
		-0.6 & \leq Q \leq 0.44.
	\end{align}
\end{subequations}
The optimal solution of Problem \eqref{VVCexample} lies exactly at breakpoint $U_4$, which hinders \textsc{Ipopt}'s ability to reduce dual infeasibility and causes it to diverge.
However, replacing \eqref{VVC_constraint} with the \softplus{} approximation \eqref{softplus_VVC} or the \spline{} approximation \eqref{eq_Qspline} enables \textsc{Ipopt} to converge for any $\epsilon \in \{1,10^{-1},10^{-2},10^{-3},10^{-4},10^{-5}\}$.\footnote{\textsc{Ipopt} fails to converge for $\epsilon \in \{10^{-6},10^{-7},10^{-8}\}$.}
Consequently, to avoid potential divergence due to nondifferentiability, a prudent choice for ensuring maximum reliability while maintaining high accuracy is \softplus{} or \spline{} with $\epsilon = 10^{-5}$.
This is also evidenced in Fig. \ref{fig_comprehensivecomparison}, which shows that \softplus{} with $\epsilon = 10^{-5}$ consistently outperforms \spline{}, producing solutions with ${\rm RelGaps}$ below $10^{-6}$\% and maximum VVWC errors on the order of $10^{-4}$\%, well below the $\pm 7.5\%$ voltage accuracy bounds of real-world smart inverters.

Finally, an interesting yet expected outcome is the difference in objective function values, which are consistently equal to or higher in Table \ref{tab:539bus_comparison_with_opendss_optimality} than in Table \ref{tab:539bus_high_comparison_objective}.
This is mathematically consistent, as the addition of a binding constraint to any optimization problem necessarily increases the objective function value. Nonetheless, the introduction of limits on phase and neutral voltages reduces the voltage unbalance factor (VUF) during solar hours.

\begin{table*}[t!]
	\centering
	\caption{Comparison of \nonsmooth{}, \softplus{}, and \spline{} approximations across all time steps on the \largenw{} system under the High PV adoption scenario, with all inequality constraints enforced and phase-to-neutral voltages as inputs to the VVWC (MA27, $\epsilon =10^{-8}$, max. ${\rm RelGap}$ = 2.65E-08\%).}
	\begin{tabular}{r rrr rrr c c}
		\toprule
		\textbf{Time} &
		\multicolumn{3}{c}{\textbf{Objective value (kW)}} &
		\multicolumn{3}{c}{\textbf{Max VVWC error (\%)}} &
		\textbf{Dispatched PV (kW)} &
		\textbf{VUF (\%)} \\
		\cmidrule(l){2-4} \cmidrule(l){5-7}
		& \nonsmooth & \softplus{} & \spline{}
		& \nonsmooth & \softplus{} & \spline{}
		& \nonsmooth & \nonsmooth \\
		\midrule
		08:00 & 165.89  & 165.89  & 165.89  & 5.66E-13 & 5.26E-13 & 5.66E-13 & 0.00    & 0.86 \\
		08:30 & -60.53  & -60.53  & -60.53  & 4.90E-12 & 4.90E-12 & 4.90E-12 & 206.36  & 0.57 \\
		09:00 & -273.44 & -273.44 & -273.44 & 4.60E-11 & 4.60E-11 & 4.60E-11 & 409.23  & 1.90 \\
		09:30 & -458.82 & -458.82 & -458.82 & 3.50E-11 & 3.48E-11 & 3.50E-11 & 604.70  & 4.01 \\
		10:00 & -612.24 & -612.24 & -612.24 & 6.64E-13 & 6.81E-13 & 6.64E-13 & 782.40  & 4.48 \\
		10:30 & -723.67 & -723.67 & -723.67 & 9.25E-08 & 9.25E-08 & 9.25E-08 & 916.91  & 4.27 \\
		11:00 & -803.75 & -803.75 & -803.75 & 8.95E-08 & 8.95E-08 & 8.95E-08 & 1013.96 & 3.99 \\
		11:30 & -861.59 & -861.59 & -861.59 & 1.13E-07 & 1.13E-07 & 1.13E-07 & 1088.85 & 3.71 \\
		12:00 & -902.46 & -902.46 & -902.46 & 1.06E-07 & 1.06E-07 & 1.06E-07 & 1139.89 & 3.61 \\
		12:30 & -932.76 & -932.76 & -932.76 & 8.84E-08 & 8.84E-08 & 8.84E-08 & 1179.44 & 3.52 \\
		13:00 & -947.37 & -947.37 & -947.37 & 7.75E-08 & 7.75E-08 & 7.75E-08 & 1199.94 & 3.51 \\
		13:30 & -952.74 & -952.74 & -952.74 & 7.58E-08 & 7.58E-08 & 7.58E-08 & 1209.21 & 3.52 \\
		14:00 & -954.06 & -954.06 & -954.06 & 7.61E-08 & 7.61E-08 & 7.61E-08 & 1214.50 & 3.50 \\
		14:30 & -948.36 & -948.36 & -948.36 & 7.70E-08 & 7.70E-08 & 7.70E-08 & 1215.86 & 3.55 \\
		15:00 & -935.44 & -935.44 & -935.44 & 1.22E-07 & 1.22E-07 & 1.22E-07 & 1210.85 & 3.58 \\
		15:30 & -915.10 & -915.10 & -915.10 & 8.44E-08 & 8.44E-08 & 8.44E-08 & 1203.06 & 3.68 \\
		16:00 & -873.92 & -873.92 & -873.92 & 9.37E-08 & 9.37E-08 & 9.37E-08 & 1173.53 & 3.78 \\
		16:30 & -818.35 & -818.35 & -818.35 & 8.33E-08 & 8.33E-08 & 8.33E-08 & 1128.96 & 4.14 \\
		17:00 & -738.35 & -738.35 & -738.35 & 7.52E-08 & 7.52E-08 & 7.52E-08 & 1054.47 & 4.56 \\
		17:30 & -632.39 & -632.39 & -632.39 & 1.03E-07 & 1.03E-07 & 1.03E-07 & 947.38  & 4.80 \\
		18:00 & -490.64 & -490.64 & -490.64 & 3.33E-11 & 3.33E-11 & 3.33E-11 & 789.79  & 4.45 \\
		18:30 & -325.78 & -325.78 & -325.78 & 4.63E-11 & 4.64E-11 & 4.63E-11 & 604.79  & 2.77 \\
		19:00 & -140.43 & -140.43 & -140.43 & 5.28E-11 & 5.28E-11 & 5.28E-11 & 408.88  & 1.05 \\
		19:30 & 46.82   & 46.82   & 46.82   & 5.44E-12 & 5.44E-12 & 5.44E-12 & 206.23  & 0.78 \\
		20:00 & 241.99  & 241.99  & 241.99  & 5.43E-13 & 5.23E-13 & 5.43E-13 & 0.00    & 1.89 \\
		\bottomrule
	\end{tabular}
	\label{tab:539bus_high_comparison_objective}
\end{table*}

\begin{table*}[t!]
	\centering
	\caption{Computation time and iteration counts for \nonsmooth{}, \softplus{}, and \spline{} formulations across all time steps on the \largenw{} system under the High PV adoption scenario, with all inequality constraints enforced and phase-to-neutral voltages as inputs to the VVWC (MA27, $\epsilon =10^{-8}$).}
	\begin{tabular}{r rrr rrr}
		\toprule
		\textbf{Time} &
		\multicolumn{3}{c}{\textbf{Computation time (s)}} &
		\multicolumn{3}{c}{\textbf{Iterations (-)}} \\
		\cmidrule(l){2-4} \cmidrule(l){5-7}
		& \nonsmooth & \softplus{} & \spline{}
		& \nonsmooth & \softplus{} & \spline{} \\
		\midrule
		08:00 & 0.87 & 0.81 & 0.87 & 20 & 20 & 20 \\
		08:30 & 0.45 & 0.47 & 0.49 & 10 & 10 & 10 \\
		09:00 & 0.53 & 0.53 & 0.57 & 12 & 12 & 12 \\
		09:30 & 0.59 & 0.62 & 0.67 & 14 & 14 & 14 \\
		10:00 & 1.15 & 1.12 & 1.14 & 28 & 28 & 28 \\
		10:30 & 1.28 & 1.25 & 1.32 & 34 & 34 & 34 \\
		11:00 & 1.29 & 1.43 & 1.32 & 34 & 34 & 34 \\
		11:30 & 1.33 & 1.32 & 1.39 & 37 & 37 & 37 \\
		12:00 & 1.47 & 1.48 & 1.63 & 40 & 40 & 40 \\
		12:30 & 1.40 & 1.52 & 1.69 & 38 & 38 & 38 \\
		13:00 & 1.38 & 1.48 & 1.54 & 39 & 39 & 39 \\
		13:30 & 1.64 & 1.75 & 1.70 & 42 & 42 & 42 \\
		14:00 & 1.45 & 1.53 & 1.58 & 39 & 39 & 39 \\
		14:30 & 1.47 & 1.49 & 1.59 & 38 & 38 & 38 \\
		15:00 & 1.34 & 1.47 & 1.55 & 37 & 37 & 37 \\
		15:30 & 1.41 & 1.42 & 1.49 & 39 & 39 & 39 \\
		16:00 & 1.52 & 1.48 & 1.63 & 42 & 42 & 42 \\
		16:30 & 1.28 & 1.30 & 1.34 & 35 & 35 & 35 \\
		17:00 & 1.52 & 1.52 & 1.63 & 41 & 41 & 41 \\
		17:30 & 1.19 & 1.19 & 1.25 & 32 & 32 & 32 \\
		18:00 & 0.69 & 0.67 & 0.70 & 16 & 16 & 16 \\
		18:30 & 0.59 & 0.59 & 0.64 & 14 & 14 & 14 \\
		19:00 & 0.54 & 0.55 & 0.57 & 12 & 12 & 12 \\
		19:30 & 0.44 & 0.48 & 0.50 & 10 & 10 & 10 \\
		20:00 & 0.73 & 0.73 & 0.77 & 18 & 18 & 18 \\
		\bottomrule
	\end{tabular}
	\label{tab:539bus_high_comparison_solvetime}
\end{table*}

\section{Discussion}\label{sec_discussion}

Gradient-based nonlinear programming approaches such as the PDIPM, with implementations available in \textsc{Knitro} \cite{KNITRO}, \textsc{Ipopt} \cite{ipopt}, and \textsc{Mips} (\textsc{Matpower}) \cite{matpowermips} are gaining considerable attention due to their tractability and increased reliability for solving large-scale nonconvex optimization problems in power systems.
This family of algorithms inherits certain properties from Newton's method, for instance requiring the (nonlinear) functions to be twice continuously differentiable.
Luckily, the alternating current (AC) generalization of Kirchhoff's circuit laws, Ohm's law, and the definition of complex power are twice continuously differentiable (and therefore smooth) functions, which can be written as a set of nonconvex quadratic constraints.
Recent implementations of these methods in the ARPA-E grid optimization competition, where the majority of leading teams used this family of PDIPMs, demonstrated their reliability and scalability.
Similar findings were observed on the four-wire OPF in unbalanced LV networks, which was recently demonstrated in \cite{Claeys2022_Fourwire} to have solve times of around one second on cases with up to 539 buses $\times$ 4 nodes/bus when using the PDIPM implementation in \textsc{Ipopt} \cite{ipopt}.
Although PDIPMs on nonconvex problems only certify local optimality, as opposed to global optimality \cite{Nocedal2006_NumericalOptimization}, a major advantage of using nonconvex models is the  feasibility of the solution (when they converge), i.e., no further projection (feasibility recovery) or validation is required to implement the optimized set points in a real-world network control system.

However, the differentiability requirement introduces more challenges when \emph{piecewise-linear} (PWL) functions such as VVC and VWC appear as constraints in optimization problems.
Because they are nondifferentiable at their breakpoints, PWL functions can introduce numerical issues in Newton-type methods and PDIPMs, potentially impeding convergence.
Within the community of practitioners, it is generally understood that nondifferentiability does not always spell convergence issues, but it is often a contributing factor.%

Nonetheless, state-of-the-art PDIPM solvers such as IPOPT are well-suited to problems that include PWL droop control functions.
Their line-search procedures adaptively reduce step sizes when encountering abrupt changes in gradients, allowing the algorithm to approach nondifferentiable regions gradually instead of stepping across them.
When the Jacobian changes sharply due to a transition between PWL segments, the solver backtracks, recomputes a search direction, and updates its barrier parameters to maintain numerical stability.
This adaptive behavior mitigates the divergence and oscillations often observed in incremental control methods and Newton-type methods without adaptive line-search.

In addition to a comprehensive benchmarking campaign across a wide range of operating conditions, feeder topologies and size, PV adoption levels, and VVWC settings, the paper examines several linear solvers available through HSL (MA27, MA57, MA77, MA86, MA97) and MUMPS, all of which have distinct factorization strategies, pivoting rules, and memory characteristics.
These solvers are known to behave differently on indefinite or ill-conditioned KKT systems arising in PDIPMs.
Expansive analysis shows that MA27 offers the best tradeoff between robustness and runtime for the types of KKT systems produced by the VVWO formulations.
This observation is consistent with the well-documented strengths of MA27: its stable symmetric indefinite factorization, predictable memory behavior, and low overhead for medium-scale sparse systems. MA57 and MA97, while more modern, exhibited slower factorization times for the problems in this work, and MUMPS showed consistently slower convergence.
Thus, the identification of MA27 as the most suitable linear solver is grounded in both systematic experimentation and the theoretical characteristics of the underlying factorization methods.
This observation is in line with others, e.g., \cite{Tasseff2019_LinearSolvers}, who have observed MA27 and MA57 being the preferred linear subsystem solvers for OPF-type problems.

Furthermore, the paper also thoroughly evaluates the impact of the smoothing parameter $\epsilon$ in the \softplus{} and \spline{}.
Small values of $\epsilon$ closely replicate the original PWL functions but could detract from reliability, whereas very large values overly smoothen these curves and decrease the approximation accuracy.
Rigorous benchmarking shows that the \softplus{} with $\epsilon =10^{-5}$ offers an optimal tradeoff between computational efficiency and approximation accuracy while ensuring high numerical stability, i.e., reliability.

\section{Conclusion}\label{sec_conclusion}

This paper proposes and compares three nonlinear models for Volt-VAr-Watt optimization (VVWO) in four-wire low-voltage (LV) networks, and identifies the best parameters, as well as the most suitable linear solvers and tools for improving numerical stability, that deliver optimal tradeoffs between reliability and computational performance.

When all inequality constraints, including voltage bounds, are removed, the three proposed nonlinear VVWO models yielded the same solutions as \textsc{OpenDSS} on the 30-bus and 99-bus systems.
However, unlike \textsc{OpenDSS}, which failed to converge in many instances of high PV generation on the 539-bus system under the Medium and High PV adoption scenarios, the three proposed nonlinear VVWO models consistently converged across all instances and all three networks under every PV adoption scenario, numerically substantiating their reliability.

When all inequality constraints were reintroduced, to demonstrate the capabilities unique to optimization-based models, two key findings were observed.
First, despite using 37.5\% fewer binary variables than standard implementations, the proposed mixed-integer nonlinear programming (MINLP) implementation was, on average, at least one order of magnitude slower than the nonlinear counterparts.
Second, a stringent choice of the approximation parameter in the smooth models produced high-quality and high-accuracy feasible solutions with comparable computational performance to the nonsmooth model.
However, to avoid potential numerical instability, this paper recommends using the softplus approximation with $\epsilon$ not exceeding $10^{-5}$.
This choice is corroborated by comprehensive numerical evaluation to offer the best compromise between reliability, computational performance, and practical accuracy, which remains well within the $\pm 7.5\%$ voltage accuracy bounds of real-world smart inverters.
In summary, this work is the first to demonstrate accurate, reliable, and computationally efficient VVWO on real four-wire unbalanced LV network models.

\ifCLASSOPTIONcaptionsoff
\newpage
\fi

\bibliographystyle{IEEEtran}
\bibliography{VVWOinFourWireLVDN}

\end{document}